\documentclass{aa}
\usepackage{graphicx}
\usepackage{natbib}
\usepackage[colorlinks=true, citecolor=blue, linkcolor=blue, urlcolor=black]{hyperref}
\usepackage[varg]{txfonts}
\usepackage{appendix}
\usepackage{comment}
\usepackage{mathrsfs}
\usepackage{amsmath}
\usepackage{multicol}
\usepackage{multirow,tabularx}
\usepackage{longtable}
\DeclareUnicodeCharacter{2212}{-}
\newcommand{\RNum}[1]{\uppercase\expandafter{\romannumeral #1\relax}}

\begin{document}
   \title{Plausible association of distant late M dwarfs with low-frequency radio emission \thanks{The spectra and photometric measurements are available in electronic form at the CDS via anonymous ftp to cdsarc.cds.unistra.fr (130.79.128.5) or via \url{https://cdsarc.cds.unistra.fr/cgi-bin/qcat?J/A+A/}} \thanks{Based on observations obtained with the Hobby-Eberly Telescope (HET), which is a joint project of the University of Texas at Austin, Pennsylvania State University, Ludwig-Maximillians-Universitaet Muenchen, and Georg-August Universitaet Goettingen. The HET is named in honor of its principal benefactors, William P. Hobby and Robert E. Eberly.}}
   \author{A. J. Gloudemans \inst{\ref{inst1}}
   \and J.~R.~Callingham \inst{\ref{inst_astron},\ref{inst1}} 
   \and K. J. Duncan \inst{\ref{inst_edin}}
   \and A. Saxena \inst{\ref{inst_ox}, \ref{inst_ucl}}
   \and Y. Harikane \inst{\ref{inst_tokyo}}
   \and G. J. Hill \inst{\ref{inst_mcdonald},\ref{inst_texas}}
   \and G. R. Zeimann \inst{\ref{inst_het}}
   \and H.~J.~A.~R\"{o}ttgering \inst{\ref{inst1}}
   \and M. J. Hardcastle \inst{\ref{inst_hertfordshire}}
   \and J. S. Pineda \inst{\ref{inst_colorado}}
   \and T. W. Shimwell \inst{\ref{inst1}, \ref{inst_astron}}
   \and D. J. B. Smith \inst{\ref{inst_hertfordshire}}
   \and J. D. Wagenveld \inst{\ref{inst_bonn}}
   }
   
   \institute{Leiden Observatory, Leiden University, PO Box 9513, 2300 RA Leiden, The Netherlands \\ e-mail: gloudemans@strw.leidenuniv.nl\label{inst1}
   \and ASTRON, Netherlands Institute for Radio Astronomy, Oude Hoogeveensedijk 4, Dwingeloo, 7991 PD, The Netherlands\label{inst_astron}
   \and Institute for Astronomy, Royal Observatory, Blackford Hill, Edinburgh, EH9 3HJ, UK\label{inst_edin} \and  Department of Physics, University of Oxford, Denys Wilkinson Building, Keble Road, Oxford OX1 3RH, UK \label{inst_ox} \and Department of Physics and Astronomy, University College London, Gower Street, London WC1E 6BT, UK \label{inst_ucl}\and Institute for Cosmic Ray Research, the University of Tokyo, 5-1-5 Kashiwa-no-Ha, Kashiwa City, Chiba, 277-8582, Japan \label{inst_tokyo} \and McDonald Observatory, University of Texas at Austin, 2515 Speedway, Stop C1402, Austin, TX 78712, USA \label{inst_mcdonald} \and Department of Astronomy, University of Texas at Austin, 2515 Speedway, Stop C1400, Austin, TX 78712, USA \label{inst_texas} \and Hobby Eberly Telescope, University of Texas, Austin, TX 78712, USA\label{inst_het} \and Centre for Astrophysics Research, University of Hertfordshire, College Lane, Hatfield AL10 9AB, UK \label{inst_hertfordshire} \and University of Colorado Boulder, Laboratory for Atmospheric and Space Physics, 3665 Discovery Drive, Boulder, CO 80303, USA\label{inst_colorado} \and Max-Planck Institut für Radioastronomie, Auf dem Hügel 69, 53121 Bonn, Germany\label{inst_bonn}}
   
   \date{Received: 09 June 2023 / Accepted: 14 August 2023}
 
 \abstract{We present the serendipitous discovery of 8 distant ($>$ 50 pc) late M dwarfs with plausible associated radio emission at 144\,MHz. The M dwarf nature of our sources has been confirmed with optical spectroscopy performed using HET/LRS2 and Subaru/FOCAS, and their radio flux densities are within the range of 0.5-1.0 mJy at 144\,MHz. Considering the radio-optical source separation and source densities of the parent catalogues, we suggest that it is statistically probable the M\,dwarfs are associated with the radio emission. However, it remains plausible that for some of the sources the radio emission originates from an optically faint and red galaxy hiding behind the M\,dwarf. The isotropic radio luminosities ($\sim10^{17-18}$ erg s$^{-1}$ Hz$^{-1}$) of the M\,dwarfs suggest that if the association is real, the radio emission is likely driven by a coherent emission process produced via plasma or electron-cyclotron maser instability processes, which is potentially caused by binary interaction. Long term monitoring in the radio and high-resolution radio follow-up observations are necessary to search for any variability and pinpoint the radio emission to determine whether our tentative conclusion that these ultracool dwarfs are radio emitting is correct. If the low-frequency radio emission is conclusively associated with the M\,dwarfs, this would reveal a new population of optically faint and distant ($>$ 50 pc) radio emitting M\,dwarfs. }

 \keywords{Stars: low-mass -- Radio continuum: stars -- Techniques: spectroscopy}  

\maketitle

\section{Introduction}
\label{sec:introduction}

M\,dwarfs are the most common stellar objects in our Milky Way \citep{Bochanski2010AJ....139.2679B}, and they exhibit strong signatures of magnetic activity all the way to the smallest masses, despite significant changes in the internal structure requiring different dynamo mechanisms from the solar case \citep[e.g.][]{reiners2010A&A...522A..13R, Cohen2015ApJ...806...41C}. The population of M\,dwarfs partly overlaps with the population of ultracool dwarfs, which have a spectral type $\geq$ M7, and consists of both very low-mass stars ($\lesssim 0.1$ M$_{\odot}$) and brown dwarfs ($\lesssim$ 0.072 M$_{\odot}$), which are unable to fuse hydrogen in their core. The properties of brown dwarfs are consistent with behaviours spanning stellar to planetary regimes. For example, such objects have been shown to exhibit optical and radio auroral emission in their stellar coronae, which can directly probe magnetospheric properties of the objects (e.g. \citealt{Hallinan2015Natur.523..568H, Pineda2017ApJ...846...75P, Kao2018ApJS..237...25K, Vedantham2020ApJ...903L..33V,Callingham2021A&A...648A..13C,2022ApJ...935...99B}). 

Because of their low luminosities, ultracool dwarfs are difficult to detect at large distances. Wide-area infrared surveys, such as the Two-Micron All-Sky Survey (2MASS; \citealt{Skrutskie2006AJ....131.1163S}), UKIRT Infrared Deep Sky Survey (UKIDSS; \citealt{Lawrence2007MNRAS.379.1599L}) and Wide-field Infrared Survey Explorer (WISE; \citealt{wright2010AJ....140.1868W}) have been crucial in uncovering ultracool dwarf populations. Infrared wide-area surveys remain the main resource for finding new cool dwarfs. However, compiling unbiased samples of these sources without spectroscopic observing campaigns remains challenging (e.g. \citealt{west2011ASPC..448.1407W, ponte2023ultracool}). Alternative search methods outside of the infrared, such as radio and X-ray observations, offer the potential for finding distinct populations of ultracool dwarfs with particular properties \citep[e.g.][]{Vedantham2020ApJ...903L..33V, DeLuca2020A&A...634L..13D}.

Radio observations of cool dwarfs can be used to study the strength and topology of their magnetic fields \citep{Berger2001Natur.410..338B, Hallinan2007ApJ...663L..25H, Route2012ApJ...747L..22R, Kao2016ApJ...818...24K}, with low-frequency radio observations ($\leq$ 200 MHz) being sensitive to the conditions of the outer corona and magnetosphere. Only recently, the first discoveries of low-frequency radio detections of M dwarfs have been reported by \cite{Lynch2017ApJ...836L..30L} using the Murchison Widefield Array (MWA; \citealt{Tingay2013PASA...30....7T}) and \cite{Vedantham2020ApJ...903L..33V} and \cite{Callingham2021NatAs...5.1233C} using the Low Frequency Array (LOFAR; \citealt{vanHaarlem2013A&A...556A...2V}). \cite{Callingham2021NatAs...5.1233C} and \cite{Vedantham2020ApJ...903L..33V} partially attribute the radio emission to plasma emission from the active chromospheres of the stars. However, the origin of radio emission for the quiescent stars in their sample is potentially driven by star-planet interactions generated via the electron-cyclotron maser instability (ECMI; \citealt{Hallinan2007ApJ...663L..25H, Route2016ApJ...830...85R, Kao2018ApJS..237...25K}), meaning low-frequency radio observations could potentially be used to find and study exoplanet magnetospheres as well. These radio processes have been seen as well in early M-types such as CR Draconis \citep{Callingham2021A&A...648A..13C} and YZ Ceti \citep{Trigilio2023arXiv230500809T}. Also, close stellar binaries, such as RS Canum Venaticorum (RS CVn) systems, have also been observed to generate radio emission via similar processes observed on chromospherically-active stars \citep[e.g.][]{2021A&A...654A..21T,2022ApJ...926L..30V}.

M\,dwarfs are well-known contaminants of optical or near-infrared high redshift quasar selection techniques because of their similar colours (e.g. \citealt{Hewett2006MNRAS.367..454H, Findlay2012MNRAS.419.3354F, Banados2016ApJS..227...11B, Wagenveld2022A&A...660A..22W}). In a campaign searching for radio-detected high-redshift quasars, \cite{Gloudemans2022A&A...668A..27G} serendipitously discovered a sample of late M\,dwarfs that were plausibly associated with low-frequency radio emission. This sample has been selected from a combination of the LOFAR Two Metre Sky Survey (LoTSS-DR2; \citealt{Shimwell2022A&A...659A...1S}) and the DESI Legacy Imaging Survey \citep{dey2019AJ....157..168D}, which is reaching over a magnitude deeper in the optical $g$-, $r$-, and $z$-band than previous all-sky optical surveys like PANSTARRS-1 (PS1; \citealt{Chambers2016arXiv161205560C}) and the Sloan Digital Sky Survey (SDSS; \citealt{York2000AJ....120.1579Y}). If the radio emission is associated with the M\,dwarfs, this would reveal a new population of distant ($>$ 50 pc) radio emitting M\,dwarfs. In this paper, we therefore present the observations and measured optical/radio properties of the M\,dwarf sample, the probability of association between the optical and radio sources, and discuss the potential origin of the radio emission. 

This paper is structured as follows. In Section \ref{sec:catalogues}, we describe the optical and radio catalogues and the cross-matching procedure to obtain the M\,dwarf sample, which we use to investigate the possibility of the radio sources being random chance associations. In Section \ref{sec:sample}, we present the optical spectra and photometry of the M\,dwarf sample, which allows for determining the spectral type and obtaining distance estimates, and we provide their measured radio properties. Finally, in Section \ref{sec:discussion} we discuss the possible physical origin of the radio emission and in Section \ref{sec:conclusion} we summarise the results and outline the future prospects. Throughout this work, we use the AB magnitude system \citep{Oke1983ApJ...266..713O}. 

\section{Description of catalogues}
\label{sec:catalogues}

The late M\,dwarfs in this work have been identified in a sample of high-$z$ quasar candidates using a Lyman break colour selection combined with a photometric redshift analysis (see \citealt{Duncan2022MNRAS.512.3662D} for more information) using the DESI Legacy Imaging Survey \citep{dey2019AJ....157..168D}. In summary, the selection required a photometric redshift $z_{\text{phot}} > 3.5 $, a LoTSS-DR2 detection within 2$\arcsec$, and a  colour redder than $\geq1.4$ magnitude between the Legacy $r$ and $z$-band. Additional spectral energy distribution (SED) fitting using the photometric redshift code \textsc{Eazy} \citep{brammer2011ApJ...739...24B} and visual inspection resulted in a sample of 142 candidates for follow-up spectroscopic observations. 

Between 2020-2022, optical spectroscopic observations have been conducted of 80 sources from this sample, using primarily the Faint Object Camera and Spectrograph (FOCAS; \citealt{Kashikawa2002PASJ...54..819K}) on the Subaru Telescope and LRS2 \citep{Chonis2016SPIE.9908E..4CC} on the Hobby Eberly Telescope (HET; \citealt{Ramsey1998AAS...193.1007R, Hill2021AJ....162..298H}). These observations lead to the discovery of 24 new quasars at $4.9<z<6.6$ and 56 other sources (see \citealt{Gloudemans2022A&A...668A..27G}). Out of these 56 other sources, we have identified 28 late M\,dwarfs. The other 56 sources are potentially low-$z$ star forming galaxies or dwarfs that are difficult to classify due to low signal-to noise. Upon further consideration of the radio-optical cross-match, we show in the following section that 8 have a significant probability of association. We note that the other 20 M\,dwarfs do not show any notable differences in their optical or radio properties, such as colour, brightness, and estimated distance (see Section~\ref{subsec:distances}). This highlights the potential of this selection method and deep surveys to uncover new populations of M\,dwarfs. However, to provide a statistically motivated clean sample (based on their positional offsets) of radio associated M\,dwarfs, we do not include them. In this section, we discuss the construction of this sample and the possibility of false chance associations between the optical and radio sources. The observational and physical properties of the M\,dwarfs are presented in Section \ref{sec:sample}. Further details of the initial target selection and data reduction are presented in \cite{Gloudemans2022A&A...668A..27G}.

\subsection{Cross-matching procedure}
\label{subsec:cross-matching}

\begin{figure}
    \centering
    \includegraphics[width=1.0\columnwidth]{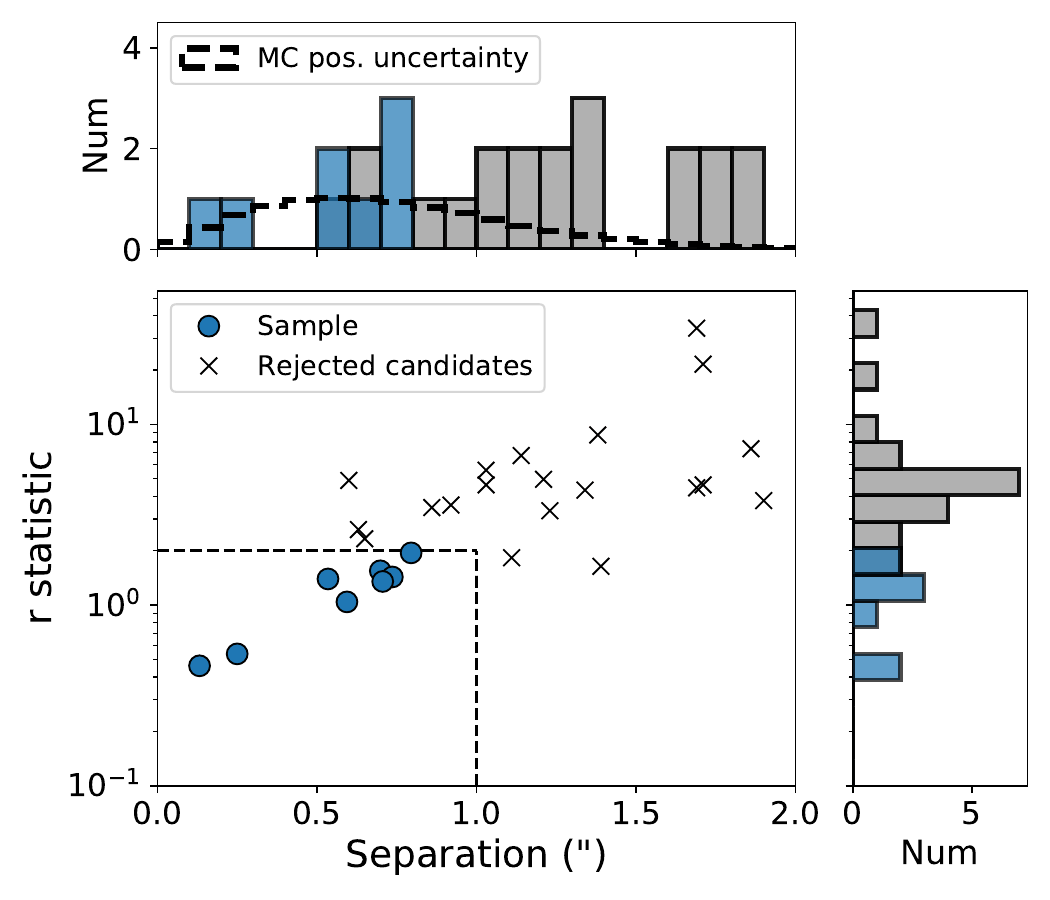}
    \caption{Radio-optical source separation of our M\,dwarf sample and rejected candidates versus the r statistic (see Eq.~\ref{eq:r_stat}). Our criteria of $r<2$ and radio-optical source separation of $<1 \arcsec$ are indicated with the black dashed line. The expected separation distribution from random sampling from the median positional uncertainties is given by the black dashed line in the top panel. The M\,dwarf sample roughly follows the expected distribution, but with a tail to larger separations.}
    \label{fig:bd_seps}
\end{figure}

As briefly mentioned before, our initial candidate sample has been obtained by cross-matching the LoTSS-DR2 catalogue ($\sim$4.4 million sources) with an optical catalogue from the Legacy surveys ($\sim$1.6 million sources) using the nearest-neighbour algorithm with a 2$\arcsec$ radius. To calculate the probability of each radio source being the true counterpart of the optical source, we use the likelihood ratio with an important parameter being the normalised radio-optical position difference given by the $r$ statistic \citep{Sutherland1992MNRAS.259..413S}, which can be explained as an uncertainty normalised angular separation metric (similar to a 2D Gaussian probability space). The $r$ value is given by

\begin{equation}
\label{eq:r_stat}
    r = \Big( \frac{(\alpha_{O} - \alpha_{R})^2}{\sigma_{\alpha, O}^2 + \sigma_{\alpha, R}^2} + \frac{(\delta_{O}-\delta_{R})^2}{\sigma_{\delta, O}^2 + \sigma_{\delta, R}^2} \Big)^{0.5},
\end{equation}

\noindent where $\alpha_{O/R}$ and $\delta_{O/R}$ are the RA and Dec. of the optical and radio source, respectively, and $\sigma_{\alpha, O/R}$ represents their individual positional uncertainties, which we assume to be Gaussian (see also \citealt{Callingham_2019}). The positional uncertainties are completely dominated by the radio -- ranging from 0.06-1.3$\arcsec$ and 0.06-0.8$\arcsec$ for the RA and Dec positions, respectively. For the $r$ value calculation we use the radio uncertainties for each source separately. For the positional uncertainty of the optical source positions in the Legacy surveys we assume 0.02$\arcsec$ \citep{dey2019AJ....157..168D}. An $r$ value of 2.0 corresponds to a reliability of $\sim$80-90\% (see \citealt{Sutherland1992MNRAS.259..413S}) and therefore we conservatively retain only those targets within $r < 2$ and radio-optical source separations of $< 1\arcsec$, as the most closely matched associations with best confidence. 

Our final sample contains 8 M\,dwarfs with their radio-optical source separations and $r$ statistic values shown in Fig.~\ref{fig:bd_seps}. For comparison we also show the rejected candidates and their $r$ values are given in Appendix \ref{sec:appendix_rejected}. The radio-optical source separations of our final sample range from 0.1-0.8$\arcsec$ and $r$ statistic values range from 0.4 to 1.9. Their median uncertainties are 0.68$\arcsec$ and 0.49$\arcsec$ for the RA and Dec positions, respectively. Sampling the radio position uncertainties in a simple Monte-Carlo simulation yields an expected offset distribution shown by the black line in the top panel of Fig.~\ref{fig:bd_seps}, when assuming no physical intrinsic offset between the radio and optical emission. This distribution is comparable to the offsets of our M\,dwarf sample and a KS-test yields a p-value of 0.16, meaning they could have been drawn from the same underlying distribution. 

We note that the offset between the radio and optical source could be partially due to proper motion. However, these M\,dwarfs are very distant (see Section~\ref{subsec:distances}) and the images are generally taken only a couple years apart. Even if these are some of the fastest moving stars (like Barnard's Star with $\sim$10$\arcsec$ yr$^{-1}$ at 1.8 pc; \citealt{Gaia2022arXiv220800211G}), it would take at least 5 years to get a 1$\arcsec$ offset due to proper motion.

\subsection{Probability of chance association}
\label{subsec:chance_assocations}

We can also estimate the chance association rate using the optical and radio source densities of the parent catalogues, which is given by  
\begin{align}
\label{eq:chance_ass}
    \mathrm{Number} \ \text{chance associations} = n_{\text{R}} \times n_{\text{O}} \times \frac{\pi \theta^2}{\Omega} , 
\end{align}
where $n_{\text{R/O}}$ are the number densities of the radio and optical sources, respectively, $\theta$ the matching radius (set to 1$\arcsec$), and $\Omega$ the overlapping survey area (5230 deg$^{2}$). Cross-matching the two parent catalogues, which are the LoTSS-DR2 and the Legacy survey catalogue with colour cuts applied, within 1$\arcsec$ results in 128 matches and this random chance association calculation suggests that 320 sources could be falsely associated, which is more than the number of matches we find. However, this calculation is based purely on the average source density, does not take into account the astrometric uncertainties of the radio and optical catalogues, and the variation in optical and radio source densities detected across the surveys, which differ significantly ($\sim100$-1000 deg$^{-2}$).

Another approach to investigate chance alignments is to perform simple Monte Carlo simulations using all the sources in the optical catalogue ($\sim$1.6 million sources), which were used for cross-matching with the LoTSS-DR2 catalogue (of $\sim$4.4 million sources). To determine the number of chance associations between these two catalogues, we determine the number densities of radio and optical sources in a $\sim$5 deg$^2$ box around each of our 128 matched candidates. Subsequently, we randomly generate the same number of optical source positions in the same sky area (1000 times) and cross-match them again to the radio sources within a 1$\arcsec$ radius. This results in a median expected number of random matches of $\sim130$ in the total LoTSS-DR2 area of 5230 deg$^2$, which is comparable to the 128 DESI-LoTSS candidates found within 1$\arcsec$. 

\begin{figure*}[ht]
    \centering
    \includegraphics[width=1.0\textwidth]{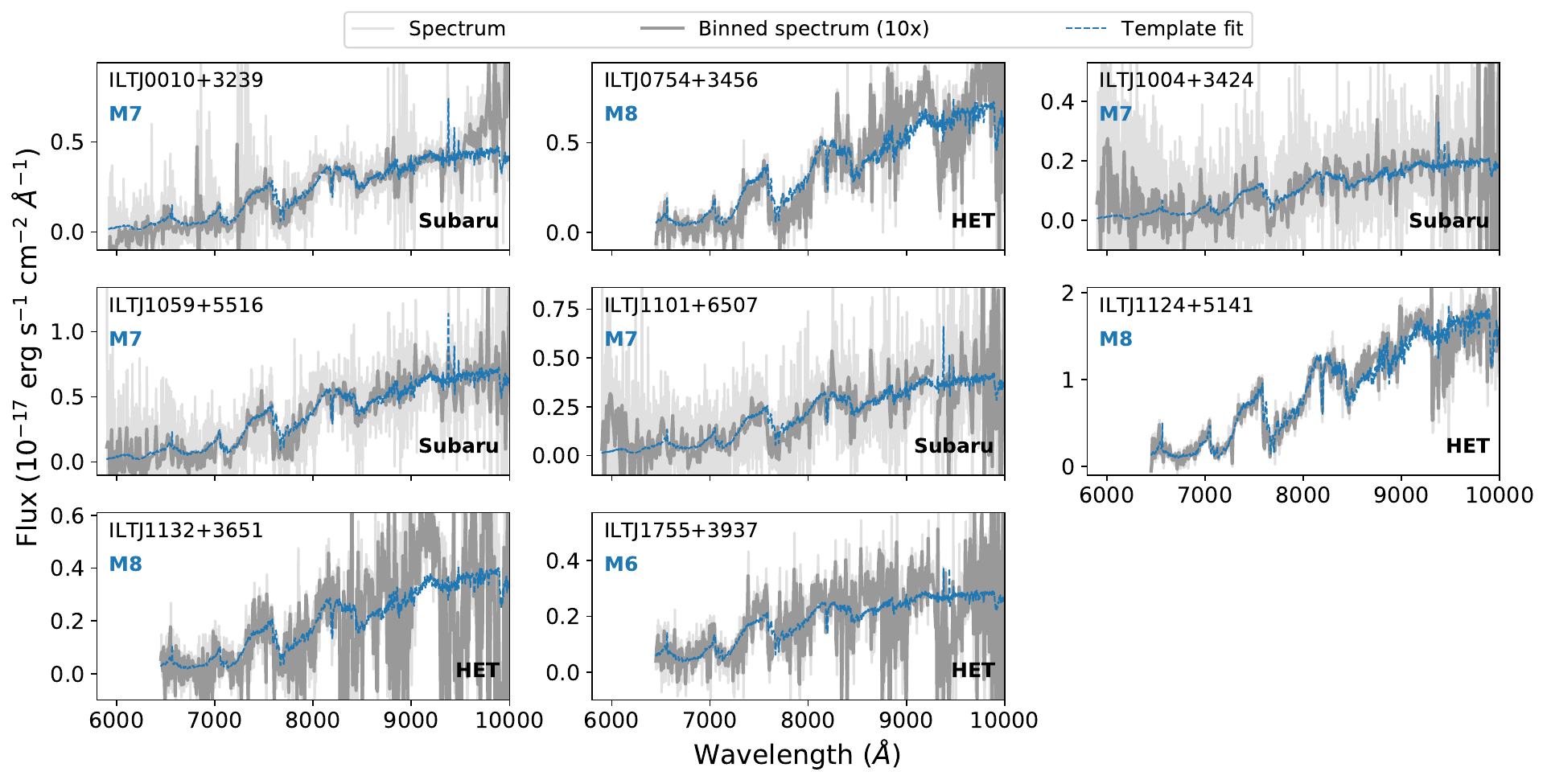}
    \caption{Observed-frame optical spectra obtained with Subaru/FOCAS and HET/LRS2 of the M\,dwarf sample. The best fitting template spectra (from PyHammer; \citealt{Roulston2020ApJS..249...34R}) are shown in blue with the M\,dwarf type indicated in the top left of each panel. We have binned each spectrum to reduce the noise level and enable a qualitative assessment of the template fits. All spectra are publicly available via anonymous ftp to cdsarc.cds.unistra.fr (130.79.128.5) or via \url{https://cdsarc.cds.unistra.fr/cgi-bin/qcat?J/A+A/}.}
    \label{fig:bd_spectra}
\end{figure*}

Therefore, we conclude it is possible that the detections are chance associations based purely on the source density and astrometric accuracy of the radio and optical catalogues. However, in case of false associations the only plausible explanation for this radio emission would be a radio-bright galaxy hiding behind the ultracool dwarf. Since there is no evidence for spectral signatures (e.g. emission lines/continuum) of galaxies in the M\,dwarf spectra (see Fig.~\ref{fig:bd_spectra}), a potential galaxy hiding behind the M\,dwarfs should be red and/or optically faint. We can therefore take a prior that for the associations to be based purely on chance, we need to hide a passive, red radio-bright galaxy closely aligned behind the ultracool dwarf. Given the measured radio flux densities of our sources ($<1$ mJy), we expect most of the radio emission to originate from star formation if there is a galaxy hiding behind the M\,dwarf (see \citealt{Best2023arXiv230505782B}). We discuss this probability in the next section.

\subsection{Hiding a faint galaxy?}
\label{subsec:hiding_galaxy}

To get an estimate of the number density of optically faint and radio bright galaxies that could be producing similar radio flux densities and remain non-detected in the obtained spectra, we utilise the LoTSS Deep Fields data in the ELAIS-N1 and Lockman Hole fields \citep{Kondapally2021A&A...648A...3K, tasse2021A&A...648A...1T, sabater2021A&A...648A...2S, Duncan2021A&A...648A...4D}. The radio images reach a sensitivity of 20 and 22 $\mu$Jy beam$^{-1}$ for the ELAIS-N1 and Lockman Hole fields, respectively. Multi-wavelength catalogues created by \cite{Kondapally2021A&A...648A...3K} contain PS1 photometry ($g,r,i,z,y$-band) from the Medium Deep Survey (MDS; \citealt{Chambers2016arXiv161205560C}) in ELAIS-N1 in an overlap area of 8.05 deg$^2$. The Lockman Hole field catalogue contains Spitzer Adaptation of the Red-sequence Cluster Survey (SpARCS; \citealt{Wilson2009ApJ...698.1943W}) photometry ($g,r,z$-band) in a 13.32 deg$^2$ area. The LoTSS-DR2 RMS noise is $\sim83$ $\mu$Jy beam$^{-1}$ and therefore to obtain a number density of faint radio emitting galaxies in LoTSS-DR2 we set a lower limit of 5$\times83$ $\mu$Jy radio flux for sources in the ELAIS-N1 and Lockman Hole catalogues. Furthermore, we require a $g$-band non-detection with $>24.6$ magnitude (5$\sigma$ limit from the Legacy Surveys) and $r$-, and $z$-band magnitudes of $>24.2$ and $>20.8$, which are the median magnitudes of our M\,dwarfs. This results in a total number of 1307 and 2130 galaxies in the ELAIS-N1 and Lockman Hole area, respectively, and number densities of 152 and 147 deg$^{-2}$ with the great majority of sources having a $z_{\text{phot}}\sim1$-3. 

To get an estimate of the sky densities of ultracool dwarfs (M-, L-, and T-dwarfs), we take the same approach as \cite{Wagenveld2022A&A...660A..22W} and use a galactic model derived by \cite{Chen2001ApJ...553..184C} to calculate the theoretical number densities of ultracool dwarfs in the Milky Way. To convert this to an observed number density, we utilise the magnitude ranges of different dwarf types listed by \cite{Best2018ApJS..234....1B} and integrate the spatial density over a cone, representing an observed sky area, for different magnitude bins (see \citealt{Wagenveld2022A&A...660A..22W} for further details). Since the ultracool dwarf number density is highly dependent on sky location, we calculate the local densities for each dwarf in our sample individually in a 1 deg$^2$ area and integrate the magnitude bins up to the Legacy $z$-band $5\sigma$ depth of 23.4 magnitude, which gives a median number density of 110 deg$^{-2}$.  

Finally, to determine the chance association of these two populations, we again use Eq.~\ref{eq:chance_ass}, which results in expected random chance associations of 0.0040 and 0.0039 deg$^{-2}$ for ELAIS-N1 and Lockman Hole, respectively. This result suggests that we can expect $\sim22$ ultracool dwarfs hiding a radio-bright galaxy, existing in the total LoTSS-DR2 area. The sky density of M\,dwarfs alone (without the L and T dwarfs) yields a similar expected of number of 20 M\,dwarfs hiding a radio-bright galaxy, since the M\,dwarfs are more abundant. Therefore, we can not rule out that some of our M\,dwarfs indeed have a faint radio galaxy hiding behind them. However, when performing the same SED fitting procedure with \textsc{Eazy} (using the PS1 $g, r, i, z, y-$band, UKIDSS $J$-band, and WISE $W1$ and $W2$ photometry) on 9948 known M-, L-, and T-dwarfs, we find that only $\sim$5\% results in a photometric redshift solution of $z_{\text{phot}}>3.5$. These sources are an almost equal combination of M, L, and T dwarfs, however since the dwarf sample contained only 359 T\,dwarfs (compared to 8381 M\,dwarfs and 1206 L\,dwarfs), the T dwarfs are most likely to have a high redshift solution. From this calculation, we expect that the vast majority (21/22) of these 22 ultracool dwarfs with a background radio-bright galaxy would not have made it to our final candidate sample. Also, considering that we have conducted spectroscopic observations of only $\sim$56\% of our high-$z$ quasar candidate sample, we conclude it remains plausible that (at least part of) the ultracool dwarf sample we present has radio emission associated with the star.

\section{Sample properties}
\label{sec:sample}

Our high-redshift quasar search has therefore led to the identification of a sample of 8 M\,dwarfs with plausible radio association. In this section, we present their spectral and photometric properties, which can be used to estimate their distances, and discuss their radio properties. 

\subsection{Optical spectra and photometric properties}

The spectra of the 8 late M\,dwarfs are shown in Fig.~\ref{fig:bd_spectra}. In short, the Subaru/FOCAS observations have been carried out using the VPH850 grating and SO58 filter ($5800 - 10350$ \r{A}) with a slit width of 1$\arcsec$ and spectral resolution of $R\sim600$. The HET spectra have been obtained using the LRS2 integral field spectrograph, which comprises a blue and red arm: LRS2-B (3650 - 6950 \r{A}) and LRS2-R (6450 - 10500 \r{A}). The raw LRS2 data were initially processed with \texttt{Panacea}\footnote{\url{https://github.com/grzeimann/Panacea}}, which carries out fiber extraction,  wavelength calibration, and relative fiber normalization. The absolute flux calibration comes from default response curves and measures of the mirror illumination as well as the exposure throughput from guider images. The data were then processed using \texttt{LRS2Multi}\footnote{\url{https://github.com/grzeimann/LRS2Multi}} to perform sky subtraction following the procedures described in \cite{Gloudemans2022A&A...668A..27G} as well as 1-D spectral extraction. The Subaru/FOCAS spectra have been reduced following the standard procedure using the Image Reduction and Analysis Facility software (IRAF; \citealt{Tody1986SPIE..627..733T}), which includes the standard bias subtraction, flat fielding, wavelength and flux calibration, and sky subtraction. 

\begin{figure}
    \centering
    \includegraphics[width=1.0\columnwidth]{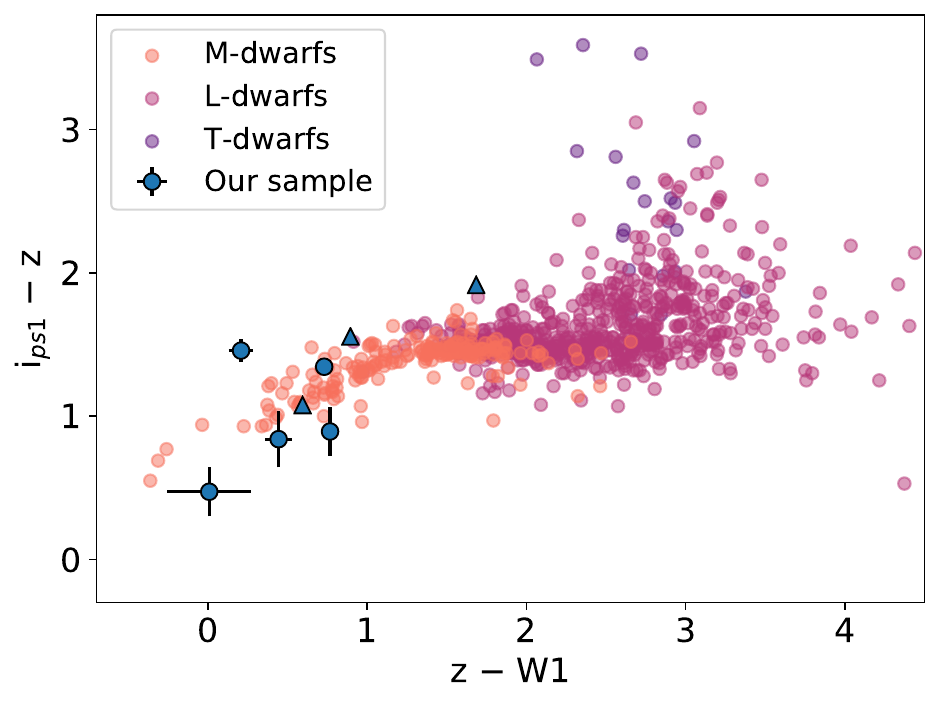}
    \caption{Optical colours of our sample compared to a literature sample of M-, L-, and T-dwarfs\textsuperscript{3}. The colours of our sample are compatible with the known M\,dwarf population. Three M\,dwarfs from our sample have a non-detection in the $i$-band and are therefore shown as lower limits.} 
    \label{fig:bd_colours}
\end{figure}

To investigate the properties of our sample, we perform stellar spectral classification on our optical spectra ($10\times$ binned to reduce the noise) using the software tool PyHammer \citep{Roulston2020ApJS..249...34R}, which automatically matches spectral types ranging from O to L type stars in the wavelength range of 3650 - 10200~\AA. To avoid the noise from telluric contamination affecting the fit, we cut-off our spectra at 9000 \AA. All our sources are classified as late M\,dwarf by PyHammer. We adjusted the subtype of 5 sources to visually better fit the spectral features. These adjustments are taken into account in the error on the spectral type, which is generally $\pm1$. The final template fits are plotted in the spectra in Fig.~\ref{fig:bd_spectra}. We do not detect significant H$\alpha$ emission lines for any of these M\,dwarfs.
However, we only have the sensitivity to detect the most active sources with these spectra, and a typical population of chromospherically active M\,dwarfs has equivalent widths (see \citealt{Reiners2010ApJ...710..924R}) that would have been within the noise. The possible origin of the radio emission will be discussed in Section \ref{subsec:physical_origin}.

The M\,dwarf nature of our sample is strengthened by their optical and near-infrared colours. In Fig.~\ref{fig:bd_colours}, we compare their colours to observed stellar M-, L-, and T-dwarfs\footnote{\url{https://docs.google.com/spreadsheets/d/1i98ft8g5mzPp2DNno0kcz4B9nzMxdpyz5UquAVhz-U8/edit?usp=sharing}} with lower limits on $i_{\text{ps}1}-z$ given for three sources that are non-detected in the PS1 $i$-band. The colours of our sample are similar to the known M\,dwarf population.

\begin{table*}
\caption{Properties of the radio-bright M\,dwarf sample. The separation reported is the offset between the optical and radio source positions from the Legacy Surveys and LoTSS-DR2 catalogue, respectively. Note that the distances we report here are estimated from the spectral types and photometry and are therefore highly uncertain. The radio luminosities are derived assuming isotropic emission.} 
\label{tab:sample_props}      
\centering
\resizebox{\textwidth}{!}{
\begin{tabular}{c c c c c c c c c}
\hline\hline
Source name & Optical coordinates  & Sep. & S$_{144MHz}$ & $z$ mag & $J$ mag & Distance & Spec. type & log$_{10}$ L$_{150\text{MHz}}$ \\ 
 & (J2000) & ($\arcsec$) & (mJy) & (AB) & (AB) & (pc) & &  (erg s$^{-1}$ Hz$^{-1})$ \\ 
\hline
ILTJ0010+3239 & 00:10:08 +32:39:01 & 0.7 & 0.97 $\pm$ 0.23 & 21.24 $\pm$ 0.04 & 20.40$\pm$0.17 & 870$\pm$600 & M7$^{+1}_{-1}$ & 17.9 \\
ILTJ0754+3456 & 07:54:22 +34:56:26 & 0.3 & 0.79 $\pm$ 0.2 & 20.32 $\pm$ 0.01 & 19.58$\pm$0.09 & 490$\pm$240 & M8$^{+1}_{-1}$ & 17.3 \\
ILTJ1004+3424 & 10:04:10 +34:24:34 & 0.7 & 0.71 $\pm$ 0.2 & 22.02 $\pm$ 0.06 & $>21.01$ & $>1200$ & M7$^{+1}_{-3}$ & - \\
ILTJ1059+5516 & 10:59:22 +55:16:21 & 0.7 & 0.62 $\pm$ 0.18 & 20.74 $\pm$ 0.02 & 20.25$\pm$0.14 & 810$\pm$550 & M7$^{+1}_{-1}$ & 17.7 \\
ILTJ1101+6507 & 11:01:31 +65:07:50 & 0.8 & 0.55 $\pm$ 0.16 & 21.18 $\pm$ 0.03 & - & - & M7$^{+1}_{-2}$ & - \\
ILTJ1124+5141 & 11:24:58 +51:41:16 & 0.6 & 0.58 $\pm$ 0.17 & 19.87 $\pm$ 0.01 & 19.25$\pm$0.07 & 420$\pm$240 & M8$^{+1}_{-1}$ & 17.1 \\
ILTJ1132+3651 & 11:32:13 +36:51:49 & 0.1 & 0.64 $\pm$ 0.15 & 20.83 $\pm$ 0.03 & 20.20$\pm$0.19 & 650$\pm$530 & M8$^{+1}_{-2}$ & 17.5 \\
ILTJ1755+3937 & 17:55:45 +39:37:44 & 0.5 & 0.92 $\pm$ 0.22 & 21.54 $\pm$ 0.04 & $>20.86$ & $>1400$ & M6$^{+2}_{-1}$ & - \\

\hline \hline
\end{tabular}}
\end{table*}

\subsection{Distances}
\label{subsec:distances}

\begin{figure}
    \centering
    \includegraphics[width=1.0\columnwidth]{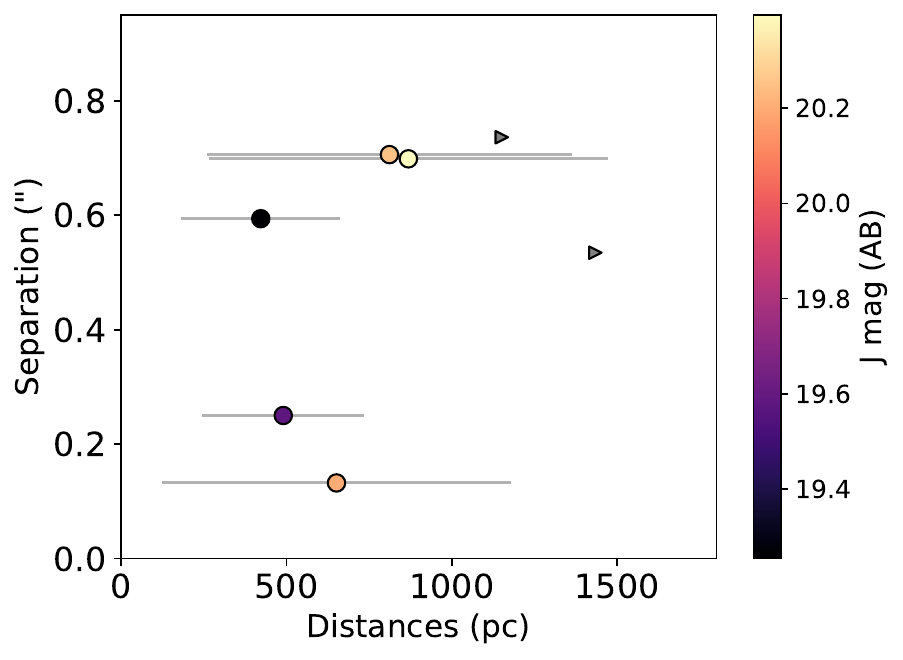}
    \caption{Estimated distances of our M\,dwarf sample versus optical-radio separation and J-band magnitude. These distances have been obtained using the relation derived by \cite{Filippazzo2015ApJ...810..158F} for ultracool dwarfs. The large error bars are mainly due to uncertainty in the spectral classification. Lower limits (3$\sigma$) on the distances are given for the two M\,dwarfs without a significant J-band detection. The M\,dwarf ILTJ1101+6507 is outside of the UHS footprint and therefore not included in this figure.}
    \label{fig:bd_distances}
\end{figure}

To estimate the distances to the M\,dwarfs in our sample, we use the absolute $J$-band magnitude versus spectral type relation derived by \cite{Filippazzo2015ApJ...810..158F}. This relation is derived for ultracool dwarfs (M6-T9) and is obtained using the optical and infrared spectra of known dwarfs together with their parallaxes and kinematic distances. We obtain the apparent $J$-band magnitudes of our sources from the UKIDSS Hemisphere Survey (UHS; \citealt{Dye2018MNRAS.473.5113D}), which range from 19.3-20.4. Two of our sources (ILTJ1004+3424 and ILTJ1755+3937) were not detected in the UHS survey with a 3$\sigma$ $J$-band detection limit of 21.01 and 20.86, respectively, obtained from forced photometry with a 1.5$\arcsec$ aperture radius. One source (ILTJ1101+6507) is not available in the UHS footprint and therefore no distance estimate is given. The resulting distances of 420-870 pc are shown in Fig.~\ref{fig:bd_distances} with lower limits for the non-detected M\,dwarfs. The error bars give the range of distances consistent with our knowledge of their spectral types and $J$ magnitude errors. Their large sizes are mainly caused by the uncertainty in the spectral type. Due to the faint nature of our sources, the distances to these sources are likely significantly higher than the distances found by \cite{Callingham2021NatAs...5.1233C} of 4 to $\sim$50 pc. We note that if these sources are equal mass binaries (see Section~\ref{sec:discussion}, we could be underestimating the distances and the distances reported here would in that case provide a lower limit.

The observed and measured properties of our M\,dwarf sample are summarised in Table ~\ref{tab:sample_props}. All the spectra, and optical and radio photometric measurements are publicly available \footnote{\url{https://cdsarc.cds.unistra.fr/cgi-bin/qcat?J/A+A/}}.

\subsection{Radio properties}
\label{subsec:radio_properties}

Since our initial sample has been selected using LOFAR, all sources in our M\,dwarf sample have radio detections at 144 MHz from the LoTSS-DR2 survey with total fluxes within the range of $\sim$0.5-1.0 mJy and estimated 144\,MHz spectral luminosities of $\sim$1-9$\times$10$^{17}$ erg Hz$^{-1}$ s$^{-1}$ (assuming isotropic emission, see Tab.~\ref{tab:sample_props}). These radio luminosities are comparable to the radio luminosity observed by \cite{Callingham2021A&A...648A..13C} of a flaring M\,dwarf binary CR Draconis. However, note that the uncertainties on the radio-luminosity are more than 100\%, since the distances are highly uncertain. 

None of our M\,dwarfs have been detected in either the VLA FIRST survey at 1.4 GHz (\citealt{Becker1994ASPC...61..165B}; 5/8 in footprint), NRAO VLA Sky Survey (NVSS; \citealt{Condon1998AJ....115.1693C}; 8/8 in footprint), or the Very Large Array Sky Survey (VLASS; \citealt{Lacy2020PASP..132c5001L}; 8/8 in footprint) at 2-4 GHz. Their optical/infrared images with radio contours are shown in Appendix \ref{sec:appendix_cutouts}. The M\,dwarfs are also not detected in the circularly polarised Stokes-V images of the LoTSS-DR2 survey (V-LoTSS; see \citealt{Callingham2023A&A...670A.124C}), unlike the M\,dwarfs discovered by \cite{Callingham2021NatAs...5.1233C}. However, it is not surprising these have not been detected in V-LoTSS, since even the total intensities (Stokes-I) fluxes are close to the noise limit of the survey (see Table ~\ref{tab:sample_props}), therefore only nearly 100\% circular polarised sources would have been detected. The median circularly polarised fraction of the M dwarf population identified by \cite{Callingham2021NatAs...5.1233C} is $\approx$60\%.

Furthermore, we investigated potential source variability by examining the single epoch LoTSS images (2-6 epochs per source). If no variability is observed, it would imply that the origin of the emission is more consistent with an extragalatic origin. However, since all of our sources are close to the detection limit of the mosaic ($\sim$0.4 mJy beam$^{-1}$ for a 5$\sigma$ detection), none are detected (signal-to-noise < 3) in the single epoch images (with $\sim$2.5-7.5 mJy beam$^{-1}$ for a 5$\sigma$ detection).

\section{Discussion}
\label{sec:discussion}

We have presented an M\,dwarf sample with plausible associated radio emission. If the low-frequency radio emission is conclusively associated with the M\,dwarfs, this would reveal a new population of radio emitting M\,dwarfs, since these are $\sim 6$-12 magnitudes fainter in the optical bands than the M\,dwarfs found by \cite{Callingham2021NatAs...5.1233C}. Also if this is the case, we expect there must be a much larger population, since less than 5\% of the known M\,dwarfs survived our photometric redshift cut (see Section \ref{subsec:hiding_galaxy}). In this section, we discuss the possible physical origin of this radio emission.

\subsection{Physical origin of radio emission}
\label{subsec:physical_origin}

Radio emission from ultracool dwarfs can be produced by either coherent or incoherent processes \citep{dulk1985}. Incoherent radio emission is generated by the gyrosynchrotron process, resulting in low ($\lesssim$20\%) circularly-polarised light and brightness temperatures $\lesssim 10^{12}$\,K. In comparison, coherent radio emission can be highly circularly polarised and can reach much higher brightness temperatures as it is produced via plasma or electron-cyclotron maser instability processes. 

The non-detection of our sample in circular polarisation implies that the circularly polarised fraction of all the sources are $<$70\%. Such a limit is not constraining for differentiating between the different emission mechanisms, especially since the median circularly polarised fraction of the nearby M\,dwarf population identified by \citet{Callingham2021NatAs...5.1233C} was $\approx$60\%. However, taking the reported distances at face value, the luminosity and implied brightness temperatures of the sample are high. For example, the reported 144\,MHz luminosities reach similar values as observed from the M\,dwarf system CR Draconis during its bursting phase \citep{Callingham2021A&A...648A..13C}. If we assume the entire photosphere of the stars are emitting, which itself is a conservative estimate since the emission regions are likely smaller \citep{zarka1998}, the brightness temperatures are $\gtrsim 10^{13}$\,K. This implies a coherent emission process is likely driving the radio emission. However, we note that the large uncertainties on the size of the ECMI emission region and distances to the sources does allow for a small parameter space in which incoherent mechanism could produce the radio emission. Follow-up observations to reduce the uncertainty on the distance to the objects would be useful to conclusively determine the emission mechanism.

We note that the properties of our population of late M dwarfs appear peculiar, in terms of their optical faintness and large distances, in light of what has been studied before for radio detected ultracool dwarfs \citep[e.g.][]{2019ApJ...871..214V,2005ApJ...621..398O}. However, there is some precedence that our low-frequency selection from a wide field survey selects for populations of objects not previously isolated. For example, \citet{Vedantham2020ApJ...903L..33V} identified a late T-dwarf object at $\approx$70\,pc, making it the most distant ultracool dwarf detected with radio emission. The radio luminosity is over an order of magnitude brighter than the ultracool dwarfs previously detected in the radio \citep[e.g.][]{Kao2016ApJ...818...24K,2019MNRAS.483..614Z}. Because of this high radio luminosity, it is unlikely for the radio emission to be driven by potential star-planet interaction, which was suggested by \cite{Callingham2021NatAs...5.1233C} to be a possible explanation for some of their radio loud M-dwarfs. This is because the suggested luminosity is too high for standard sub-Alfv\'{e}nic star-planet interaction models. 

One hypothesis for this increased radio luminosity is that the object identified by \citet{Vedantham2020ApJ...903L..33V} is a magnetically-interacting binary, which allows the radio emission to boost to the observed luminosities. It is possible that the ultracool dwarfs we have detected are actually similar mass binaries, helping explain why the radio luminosity can reach such high levels, keeping in mind the significant uncertainty of the distance of these objects. In such a case, the hypothesis is that closely interacting binaries, such as RS CVn systems, are generating the coherent radio emission via similar processes observed on chromospherically-active stars, which can produce high-brightness temperature radio emission that is relatively constant in time \citep[e.g][]{2021A&A...654A..21T,2022ApJ...926L..30V}. Similarly, \cite{Vedantham2023A&A...675L...6V} recently discovered radio pulsations from a T dwarf binary with the radio emission potentially powered by this dwarf binary interaction. Their work also highlights the capabilities of new low-frequency radio surveys in discovering radio-loud objects at planetary mass-scales.

\section{Summary and conclusion}
\label{sec:conclusion}

In a campaign searching for high-$z$ quasars, we serendipitously discovered 8 M\,dwarfs with potentially associated low-frequency radio emission at 144\,MHz. The optical spectra taken with Subaru/FOCAS and HET/LRS2 reveal they are likely late M\,dwarfs. These M\,dwarfs are optically faint ($\sim$20 mag in $J$-band) from which we have estimated distances of 420-870 pc using a spectral type distance relation. Their radio luminosities of $\sim10^{17-18}$ erg s$^{-1}$ Hz$^{-1}$ are comparable to previously observed radio stellar flares. However, since we do not see any variability in the single epoch images, the radio emission is likely emitted over a long period of time than a single 8\,h epoch. To investigate the association between the optical and radio source, we calculate the random chance association rate analytically and using Monte Carlo simulations, which suggests it is possible these are chance associations based purely on the source density and astrometric accuracy of the radio and optical catalogues. However, the only other plausible origin of the radio emission is from an optically faint and radio-bright galaxy hiding behind the M dwarf. Our ultracool dwarf and radio galaxy sky density calculations using the LOFAR Deep Fields suggest it remains plausible (at least part of) our ultracool dwarf sample has true radio emission associated to it. 

The radio luminosities of our sample suggests the radio emission is possibly driven by a coherent emission process, which can produce much higher brightness temperatures than incoherent emission. Potentially, these objects are similar mass binaries, which allows to boost the emission to the observed luminosities. This sample remains peculiar (in terms of their optical faintness and distance) in light of previously discovered ultracool dwarfs and, if the radio emission is proven to be associated, this would reveal a new population of distant late M\,dwarfs. This work highlights the utility of wide-area surveys and knowledge exchange between different fields within astronomy for serendipitous discoveries. Long term monitoring of these objects in the radio to search for any variability or proper motion is required to determine whether our tentative conclusion that these ultracool dwarfs are radio emitting is correct.

\begin{acknowledgements}
{We would like to thank Prof. S. Mahadevan for a fruitful discussion that helped guide the focus of this paper.

We are grateful for the support of Kentaro Aoki and Ichi Tanaka during our Subaru/FOCAS observations (proposal ID S21B-003).

The Low Resolution Spectrograph 2 (LRS2) was developed and funded by the University of Texas at Austin McDonald Observatory and Department of Astronomy, and by Pennsylvania State University. We thank the Leibniz-Institut fur Astrophysik Potsdam (AIP) and the Institut fur Astrophysik Goettingen (IAG) for their contributions to the construction of the integral field units.

We thank the Board of the Hobby-Eberly Telescope for the allocation of Guaranteed Time for the LRS2 instrument, which was important in enabling this investigation.

This paper is based (in part) on data obtained with the International LOFAR Telescope (ILT) under project codes LC0 015, LC2 024, LC2 038, LC3 008, LC4 008, LC4 034 and LT10 01. LOFAR \citep{vanHaarlem2013A&A...556A...2V} is the Low Frequency Array designed and constructed by ASTRON. It has observing, data processing, and data storage facilities in several countries, which are owned by various parties (each with their own funding sources), and which are collectively operated by the ILT foundation under a joint scientific policy. The ILT resources have benefited from the following recent major funding sources: CNRS-INSU, Observatoire de Paris and Universit\'e d'Orl\'eans, France; BMBF, MIWF-NRW, MPG, Germany; Science Foundation Ireland (SFI), Department of Business, Enterprise and Innovation (DBEI), Ireland; NWO, The Netherlands; The Science and Technology Facilities Council, UK; Ministry of Science and Higher Education, Poland.

This research made use of the Dutch national e-infrastructure with support of the SURF Cooperative (e-infra 180169) and the LOFAR e-infra group. The J\"ulich LOFAR Long Term Archive and the German LOFAR network are both coordinated and operated by the J\"ulich Supercomputing Centre (JSC), and computing resources on the supercomputer JUWELS at JSC were provided by the Gauss Centre for Supercomputing e.V. (grant CHTB00) through the John von Neumann Institute for Computing (NIC).

This research made use of the University of Hertfordshire high-performance computing facility and the LOFAR-UK computing facility located at the University of Hertfordshire and supported by STFC [ST/P000096/1], and of the Italian LOFAR IT computing infrastructure supported and operated by INAF, and by the Physics Department of Turin university (under an agreement with Consorzio Interuniversitario per la Fisica Spaziale) at the C3S Supercomputing Centre, Italy.

}
	
\end{acknowledgements}

\bibliographystyle{aa}
\bibliography{bibliography.bib}

\begin{thebibliography}{70}
\expandafter\ifx\csname natexlab\endcsname\relax\def\natexlab#1{#1}\fi

\bibitem[{{Ba{\~n}ados} {et~al.}(2016){Ba{\~n}ados}, {Venemans}, {Decarli},
  {Farina}, {Mazzucchelli}, {Walter}, {Fan}, {Stern}, {Schlafly}, {Chambers},
  {Rix}, {Jiang}, {McGreer}, {Simcoe}, {Wang}, {Yang}, {Morganson}, {De Rosa},
  {Greiner}, {Balokovi{\'c}}, {Burgett}, {Cooper}, {Draper}, {Flewelling},
  {Hodapp}, {Jun}, {Kaiser}, {Kudritzki}, {Magnier}, {Metcalfe}, {Miller},
  {Schindler}, {Tonry}, {Wainscoat}, {Waters}, \&
  {Yang}}]{Banados2016ApJS..227...11B}
{Ba{\~n}ados}, E., {Venemans}, B.~P., {Decarli}, R., {et~al.} 2016, \apjs, 227,
  11

\bibitem[{{Bastian} {et~al.}(2022){Bastian}, {Cotton}, \&
  {Hallinan}}]{2022ApJ...935...99B}
{Bastian}, T.~S., {Cotton}, W.~D., \& {Hallinan}, G. 2022, \apj, 935, 99

\bibitem[{{Becker} {et~al.}(1994){Becker}, {White}, \&
  {Helfand}}]{Becker1994ASPC...61..165B}
{Becker}, R.~H., {White}, R.~L., \& {Helfand}, D.~J. 1994, in Astronomical
  Society of the Pacific Conference Series, Vol.~61, Astronomical Data Analysis
  Software and Systems III, ed. D.~R. {Crabtree}, R.~J. {Hanisch}, \&
  J.~{Barnes}, 165

\bibitem[{{Berger} {et~al.}(2001){Berger}, {Ball}, {Becker}, {Clarke}, {Frail},
  {Fukuda}, {Hoffman}, {Mellon}, {Momjian}, {Murphy}, {Teng}, {Woodruff},
  {Zauderer}, \& {Zavala}}]{Berger2001Natur.410..338B}
{Berger}, E., {Ball}, S., {Becker}, K.~M., {et~al.} 2001, \nat, 410, 338

\bibitem[{{Best} {et~al.}(2023){Best}, {Kondapally}, {Williams}, {Cochrane},
  {Duncan}, {Hale}, {Haskell}, {Malek}, {McCheyne}, {Smith}, {Wang}, {Botteon},
  {Bonato}, {Bondi}, {Calistro Rivera}, {Gao}, {Gurkan}, {Hardcastle},
  {Jarvis}, {Mingo}, {Miraghaei}, {Morabito}, {Nisbet}, {Prandoni},
  {Rottgering}, {Sabater}, {Shimwell}, {Tasse}, \& {van
  Weeren}}]{Best2023arXiv230505782B}
{Best}, P.~N., {Kondapally}, R., {Williams}, W.~L., {et~al.} 2023, arXiv
  e-prints, arXiv:2305.05782

\bibitem[{{Best} {et~al.}(2018){Best}, {Magnier}, {Liu}, {Aller}, {Zhang},
  {Burgett}, {Chambers}, {Draper}, {Flewelling}, {Kaiser}, {Kudritzki},
  {Metcalfe}, {Tonry}, {Wainscoat}, \& {Waters}}]{Best2018ApJS..234....1B}
{Best}, W. M.~J., {Magnier}, E.~A., {Liu}, M.~C., {et~al.} 2018, \apjs, 234, 1

\bibitem[{{Bochanski} {et~al.}(2010){Bochanski}, {Hawley}, {Covey}, {West},
  {Reid}, {Golimowski}, \& {Ivezi{\'c}}}]{Bochanski2010AJ....139.2679B}
{Bochanski}, J.~J., {Hawley}, S.~L., {Covey}, K.~R., {et~al.} 2010, \aj, 139,
  2679

\bibitem[{{Brammer} {et~al.}(2011){Brammer}, {Whitaker}, {van Dokkum},
  {Marchesini}, {Franx}, {Kriek}, {Labb{\'e}}, {Lee}, {Muzzin}, {Quadri},
  {Rudnick}, \& {Williams}}]{brammer2011ApJ...739...24B}
{Brammer}, G.~B., {Whitaker}, K.~E., {van Dokkum}, P.~G., {et~al.} 2011, \apj,
  739, 24

\bibitem[{{Callingham} {et~al.}(2021{\natexlab{a}}){Callingham}, {Pope},
  {Feinstein}, {Vedantham}, {Shimwell}, {Zarka}, {Tasse}, {Lamy}, {Veken},
  {Toet}, {Sabater}, {Best}, {van Weeren}, {R{\"o}ttgering}, \&
  {Ray}}]{Callingham2021A&A...648A..13C}
{Callingham}, J.~R., {Pope}, B.~J.~S., {Feinstein}, A.~D., {et~al.}
  2021{\natexlab{a}}, \aap, 648, A13

\bibitem[{{Callingham} {et~al.}(2023){Callingham}, {Shimwell}, {Vedantham},
  {Bassa}, {O'Sullivan}, {Yiu}, {Bloot}, {Best}, {Hardcastle}, {Haverkorn},
  {Kavanagh}, {Lamy}, {Pope}, {R{\"o}ttgering}, {Schwarz}, {Tasse}, {van
  Weeren}, {White}, {Zarka}, {Bomans}, {Bonafede}, {Bonato}, {Botteon},
  {Bruggen}, {Chy{\.z}y}, {Drabent}, {Emig}, {Gloudemans}, {G{\"u}rkan},
  {Hajduk}, {Hoang}, {Hoeft}, {Iacobelli}, {Kadler}, {Kunert-Bajraszewska},
  {Mingo}, {Morabito}, {Nair}, {P{\'e}rez-Torres}, {Ray}, {Riseley},
  {Rowlinson}, {Shulevski}, {Sweijen}, {Timmerman}, {Vaccari}, \&
  {Zheng}}]{Callingham2023A&A...670A.124C}
{Callingham}, J.~R., {Shimwell}, T.~W., {Vedantham}, H.~K., {et~al.} 2023,
  \aap, 670, A124

\bibitem[{Callingham {et~al.}(2019)Callingham, Vedantham, Pope, Shimwell, \&
  the LoTSS~team}]{Callingham_2019}
Callingham, J.~R., Vedantham, H.~K., Pope, B. J.~S., Shimwell, T.~W., \& the
  LoTSS~team. 2019, Research Notes of the AAS, 3, 37

\bibitem[{{Callingham} {et~al.}(2021{\natexlab{b}}){Callingham}, {Vedantham},
  {Shimwell}, {Pope}, {Davis}, {Best}, {Hardcastle}, {R{\"o}ttgering},
  {Sabater}, {Tasse}, {van Weeren}, {Williams}, {Zarka}, {de Gasperin}, \&
  {Drabent}}]{Callingham2021NatAs...5.1233C}
{Callingham}, J.~R., {Vedantham}, H.~K., {Shimwell}, T.~W., {et~al.}
  2021{\natexlab{b}}, Nature Astronomy, 5, 1233

\bibitem[{{Chambers} {et~al.}(2016){Chambers}, {Magnier}, {Metcalfe},
  {Flewelling}, {Huber}, {Waters}, {Denneau}, {Draper}, {Farrow}, {Finkbeiner},
  {Holmberg}, {Koppenhoefer}, {Price}, {Rest}, {Saglia}, {Schlafly}, {Smartt},
  {Sweeney}, {Wainscoat}, {Burgett}, {Chastel}, {Grav}, {Heasley}, {Hodapp},
  {Jedicke}, {Kaiser}, {Kudritzki}, {Luppino}, {Lupton}, {Monet}, {Morgan},
  {Onaka}, {Shiao}, {Stubbs}, {Tonry}, {White}, {Ba{\~n}ados}, {Bell},
  {Bender}, {Bernard}, {Boegner}, {Boffi}, {Botticella}, {Calamida},
  {Casertano}, {Chen}, {Chen}, {Cole}, {Deacon}, {Frenk}, {Fitzsimmons},
  {Gezari}, {Gibbs}, {Goessl}, {Goggia}, {Gourgue}, {Goldman}, {Grant},
  {Grebel}, {Hambly}, {Hasinger}, {Heavens}, {Heckman}, {Henderson}, {Henning},
  {Holman}, {Hopp}, {Ip}, {Isani}, {Jackson}, {Keyes}, {Koekemoer}, {Kotak},
  {Le}, {Liska}, {Long}, {Lucey}, {Liu}, {Martin}, {Masci}, {McLean}, {Mindel},
  {Misra}, {Morganson}, {Murphy}, {Obaika}, {Narayan}, {Nieto-Santisteban},
  {Norberg}, {Peacock}, {Pier}, {Postman}, {Primak}, {Rae}, {Rai}, {Riess},
  {Riffeser}, {Rix}, {R{\"o}ser}, {Russel}, {Rutz}, {Schilbach}, {Schultz},
  {Scolnic}, {Strolger}, {Szalay}, {Seitz}, {Small}, {Smith}, {Soderblom},
  {Taylor}, {Thomson}, {Taylor}, {Thakar}, {Thiel}, {Thilker}, {Unger},
  {Urata}, {Valenti}, {Wagner}, {Walder}, {Walter}, {Watters}, {Werner},
  {Wood-Vasey}, \& {Wyse}}]{Chambers2016arXiv161205560C}
{Chambers}, K.~C., {Magnier}, E.~A., {Metcalfe}, N., {et~al.} 2016, arXiv
  e-prints, arXiv:1612.05560

\bibitem[{{Chen} {et~al.}(2001){Chen}, {Stoughton}, {Smith}, {Uomoto}, {Pier},
  {Yanny}, {Ivezi{\'c}}, {York}, {Anderson}, {Annis}, {Brinkmann}, {Csabai},
  {Fukugita}, {Hindsley}, {Lupton}, {Munn}, \& {SDSS
  Collaboration}}]{Chen2001ApJ...553..184C}
{Chen}, B., {Stoughton}, C., {Smith}, J.~A., {et~al.} 2001, \apj, 553, 184

\bibitem[{{Chonis} {et~al.}(2016){Chonis}, {Hill}, {Lee}, {Tuttle}, {Vattiat},
  {Drory}, {Indahl}, {Peterson}, \& {Ramsey}}]{Chonis2016SPIE.9908E..4CC}
{Chonis}, T.~S., {Hill}, G.~J., {Lee}, H., {et~al.} 2016, in Society of
  Photo-Optical Instrumentation Engineers (SPIE) Conference Series, Vol. 9908,
  Ground-based and Airborne Instrumentation for Astronomy VI, ed. C.~J.
  {Evans}, L.~{Simard}, \& H.~{Takami}, 99084C

\bibitem[{{Cohen} {et~al.}(2015){Cohen}, {Ma}, {Drake}, {Glocer}, {Garraffo},
  {Bell}, \& {Gombosi}}]{Cohen2015ApJ...806...41C}
{Cohen}, O., {Ma}, Y., {Drake}, J.~J., {et~al.} 2015, \apj, 806, 41

\bibitem[{{Condon} {et~al.}(1998){Condon}, {Cotton}, {Greisen}, {Yin},
  {Perley}, {Taylor}, \& {Broderick}}]{Condon1998AJ....115.1693C}
{Condon}, J.~J., {Cotton}, W.~D., {Greisen}, E.~W., {et~al.} 1998, \aj, 115,
  1693

\bibitem[{{Cutri} {et~al.}(2021){Cutri}, {Wright}, {Conrow}, {Fowler},
  {Eisenhardt}, {Grillmair}, {Kirkpatrick}, {Masci}, {McCallon}, {Wheelock},
  {Fajardo-Acosta}, {Yan}, {Benford}, {Harbut}, {Jarrett}, {Lake}, {Leisawitz},
  {Ressler}, {Stanford}, {Tsai}, {Liu}, {Helou}, {Mainzer}, {Gettngs},
  {Gonzalez}, {Hoffman}, {Marsh}, {Padgett}, {Skrutskie}, {Beck}, {Papin}, \&
  {Wittman}}]{Cutri2014yCat.2328....0C}
{Cutri}, R.~M., {Wright}, E.~L., {Conrow}, T., {et~al.} 2021, VizieR Online
  Data Catalog, II/328

\bibitem[{dal Ponte {et~al.}(2023)dal Ponte, Santiago, Rosell, Paris, Pace,
  Bechtol, Abbott, Aguena, Allam, Alves, Bacon, Bertin, Bocquet, Brooks, Burke,
  Kind, Carretero, Conselice, Costanzi, Desai, Vicente, Doel, Everett, Ferrero,
  Flaugher, Frieman, García-Bellido, Gerdes, Gruendl, Gruen, Gutierrez,
  Hinton, Hollowood, James, Kuehn, Kuropatkin, Marshall, Mena-Fernández,
  Menanteau, Miquel, Ogando, Palmese, Paz-Chinchón, Pereira, Malagón, Pieres,
  Raveri, Rodriguez-Monroy, Sanchez, Scarpine, Schubnell, Sevilla-Noarbe,
  Smith, Soares-Santos, Suchyta, Swanson, Tarle, Thomas, To, \&
  Weaverdyck}]{ponte2023ultracool}
dal Ponte, M., Santiago, B., Rosell, A.~C., {et~al.} 2023, Ultracool dwarfs
  candidates based on six years of the Dark Energy Survey data

\bibitem[{{De Luca} {et~al.}(2020){De Luca}, {Stelzer}, {Burgasser},
  {Pizzocaro}, {Ranalli}, {Raetz}, {Marelli}, {Novara}, {Vignali}, {Belfiore},
  {Esposito}, {Franzetti}, {Fumana}, {Gilli}, {Salvaterra}, \&
  {Tiengo}}]{DeLuca2020A&A...634L..13D}
{De Luca}, A., {Stelzer}, B., {Burgasser}, A.~J., {et~al.} 2020, \aap, 634, L13

\bibitem[{{Dey} {et~al.}(2019){Dey}, {Schlegel}, {Lang}, {Blum}, {Burleigh},
  {Fan}, {Findlay}, {Finkbeiner}, {Herrera}, {Juneau}, {Landriau}, {Levi},
  {McGreer}, {Meisner}, {Myers}, {Moustakas}, {Nugent}, {Patej}, {Schlafly},
  {Walker}, {Valdes}, {Weaver}, {Y{\`e}che}, {Zou}, {Zhou}, {Abareshi},
  {Abbott}, {Abolfathi}, {Aguilera}, {Alam}, {Allen}, {Alvarez}, {Annis},
  {Ansarinejad}, {Aubert}, {Beechert}, {Bell}, {BenZvi}, {Beutler}, {Bielby},
  {Bolton}, {Brice{\~n}o}, {Buckley-Geer}, {Butler}, {Calamida}, {Carlberg},
  {Carter}, {Casas}, {Castander}, {Choi}, {Comparat}, {Cukanovaite}, {Delubac},
  {DeVries}, {Dey}, {Dhungana}, {Dickinson}, {Ding}, {Donaldson}, {Duan},
  {Duckworth}, {Eftekharzadeh}, {Eisenstein}, {Etourneau}, {Fagrelius},
  {Farihi}, {Fitzpatrick}, {Font-Ribera}, {Fulmer}, {G{\"a}nsicke},
  {Gaztanaga}, {George}, {Gerdes}, {Gontcho}, {Gorgoni}, {Green}, {Guy},
  {Harmer}, {Hernandez}, {Honscheid}, {Huang}, {James}, {Jannuzi}, {Jiang},
  {Joyce}, {Karcher}, {Karkar}, {Kehoe}, {Kneib}, {Kueter-Young}, {Lan},
  {Lauer}, {Le Guillou}, {Le Van Suu}, {Lee}, {Lesser}, {Perreault Levasseur},
  {Li}, {Mann}, {Marshall}, {Mart{\'\i}nez-V{\'a}zquez}, {Martini}, {du Mas des
  Bourboux}, {McManus}, {Meier}, {M{\'e}nard}, {Metcalfe},
  {Mu{\~n}oz-Guti{\'e}rrez}, {Najita}, {Napier}, {Narayan}, {Newman}, {Nie},
  {Nord}, {Norman}, {Olsen}, {Paat}, {Palanque-Delabrouille}, {Peng},
  {Poppett}, {Poremba}, {Prakash}, {Rabinowitz}, {Raichoor}, {Rezaie},
  {Robertson}, {Roe}, {Ross}, {Ross}, {Rudnick}, {Safonova}, {Saha},
  {S{\'a}nchez}, {Savary}, {Schweiker}, {Scott}, {Seo}, {Shan}, {Silva},
  {Slepian}, {Soto}, {Sprayberry}, {Staten}, {Stillman}, {Stupak}, {Summers},
  {Sien Tie}, {Tirado}, {Vargas-Maga{\~n}a}, {Vivas}, {Wechsler}, {Williams},
  {Yang}, {Yang}, {Yapici}, {Zaritsky}, {Zenteno}, {Zhang}, {Zhang}, {Zhou}, \&
  {Zhou}}]{dey2019AJ....157..168D}
{Dey}, A., {Schlegel}, D.~J., {Lang}, D., {et~al.} 2019, \aj, 157, 168

\bibitem[{{Dulk}(1985)}]{dulk1985}
{Dulk}, G.~A. 1985, \araa, 23, 169

\bibitem[{{Duncan}(2022)}]{Duncan2022MNRAS.512.3662D}
{Duncan}, K.~J. 2022, \mnras, 512, 3662

\bibitem[{{Duncan} {et~al.}(2021){Duncan}, {Kondapally}, {Brown}, {Bonato},
  {Best}, {R{\"o}ttgering}, {Bondi}, {Bowler}, {Cochrane}, {G{\"u}rkan},
  {Hardcastle}, {Jarvis}, {Kunert-Bajraszewska}, {Leslie}, {Ma{\l}ek},
  {Morabito}, {O'Sullivan}, {Prandoni}, {Sabater}, {Shimwell}, {Smith}, {Wang},
  {Wo{\l}owska}, \& {Tasse}}]{Duncan2021A&A...648A...4D}
{Duncan}, K.~J., {Kondapally}, R., {Brown}, M.~J.~I., {et~al.} 2021, \aap, 648,
  A4

\bibitem[{{Dye} {et~al.}(2018){Dye}, {Lawrence}, {Read}, {Fan}, {Kerr},
  {Varricatt}, {Furnell}, {Edge}, {Irwin}, {Hambly}, {Lucas}, {Almaini},
  {Chambers}, {Green}, {Hewett}, {Liu}, {McGreer}, {Best}, {Zhang}, {Sutorius},
  {Froebrich}, {Magnier}, {Hasinger}, {Lederer}, {Bold}, \&
  {Tedds}}]{Dye2018MNRAS.473.5113D}
{Dye}, S., {Lawrence}, A., {Read}, M.~A., {et~al.} 2018, \mnras, 473, 5113

\bibitem[{{Filippazzo} {et~al.}(2015){Filippazzo}, {Rice}, {Faherty}, {Cruz},
  {Van Gordon}, \& {Looper}}]{Filippazzo2015ApJ...810..158F}
{Filippazzo}, J.~C., {Rice}, E.~L., {Faherty}, J., {et~al.} 2015, \apj, 810,
  158

\bibitem[{{Findlay} {et~al.}(2012){Findlay}, {Sutherland}, {Venemans},
  {Reyl{\'e}}, {Robin}, {Bonfield}, {Bruce}, \&
  {Jarvis}}]{Findlay2012MNRAS.419.3354F}
{Findlay}, J.~R., {Sutherland}, W.~J., {Venemans}, B.~P., {et~al.} 2012,
  \mnras, 419, 3354

\bibitem[{{Gaia Collaboration} {et~al.}(2022){Gaia Collaboration}, {Vallenari},
  {Brown}, {Prusti}, {de Bruijne}, {Arenou}, {Babusiaux}, {Biermann},
  {Creevey}, {Ducourant}, {Evans}, {Eyer}, {Guerra}, {Hutton}, {Jordi},
  {Klioner}, {Lammers}, {Lindegren}, {Luri}, {Mignard}, {Panem}, {Pourbaix},
  {Randich}, {Sartoretti}, {Soubiran}, {Tanga}, {Walton}, {Bailer-Jones},
  {Bastian}, {Drimmel}, {Jansen}, {Katz}, {Lattanzi}, {van Leeuwen}, {Bakker},
  {Cacciari}, {Casta{\~n}eda}, {De Angeli}, {Fabricius}, {Fouesneau},
  {Fr{\'e}mat}, {Galluccio}, {Guerrier}, {Heiter}, {Masana}, {Messineo},
  {Mowlavi}, {Nicolas}, {Nienartowicz}, {Pailler}, {Panuzzo}, {Riclet}, {Roux},
  {Seabroke}, {Sordo{\o}rcit}, {Th{\'e}venin}, {Gracia-Abril}, {Portell},
  {Teyssier}, {Altmann}, {Andrae}, {Audard}, {Bellas-Velidis}, {Benson},
  {Berthier}, {Blomme}, {Burgess}, {Busonero}, {Busso}, {C{\'a}novas}, {Carry},
  {Cellino}, {Cheek}, {Clementini}, {Damerdji}, {Davidson}, {de Teodoro},
  {Nu{\~n}ez Campos}, {Delchambre}, {Dell'Oro}, {Esquej},
  {Fern{\'a}ndez-Hern{\'a}ndez}, {Fraile}, {Garabato}, {Garc{\'\i}a-Lario},
  {Gosset}, {Haigron}, {Halbwachs}, {Hambly}, {Harrison}, {Hern{\'a}ndez},
  {Hestroffer}, {Hodgkin}, {Holl}, {Jan{\ss}en}, {Jevardat de Fombelle},
  {Jordan}, {Krone-Martins}, {Lanzafame}, {L{\"o}ffler}, {Marchal}, {Marrese},
  {Moitinho}, {Muinonen}, {Osborne}, {Pancino}, {Pauwels}, {Recio-Blanco},
  {Reyl{\'e}}, {Riello}, {Rimoldini}, {Roegiers}, {Rybizki}, {Sarro}, {Siopis},
  {Smith}, {Sozzetti}, {Utrilla}, {van Leeuwen}, {Abbas}, {{\'A}brah{\'a}m},
  {Abreu Aramburu}, {Aerts}, {Aguado}, {Ajaj}, {Aldea-Montero}, {Altavilla},
  {{\'A}lvarez}, {Alves}, {Anders}, {Anderson}, {Anglada Varela}, {Antoja},
  {Baines}, {Baker}, {Balaguer-N{\'u}{\~n}ez}, {Balbinot}, {Balog}, {Barache},
  {Barbato}, {Barros}, {Barstow}, {Bartolom{\'e}}, {Bassilana}, {Bauchet},
  {Becciani}, {Bellazzini}, {Berihuete}, {Bernet}, {Bertone}, {Bianchi},
  {Binnenfeld}, {Blanco-Cuaresma}, {Blazere}, {Boch}, {Bombrun}, {Bossini},
  {Bouquillon}, {Bragaglia}, {Bramante}, {Breedt}, {Bressan}, {Brouillet},
  {Brugaletta}, {Bucciarelli}, {Burlacu}, {Butkevich}, {Buzzi}, {Caffau},
  {Cancelliere}, {Cantat-Gaudin}, {Carballo}, {Carlucci}, {Carnerero},
  {Carrasco}, {Casamiquela}, {Castellani}, {Castro-Ginard}, {Chaoul},
  {Charlot}, {Chemin}, {Chiaramida}, {Chiavassa}, {Chornay}, {Comoretto},
  {Contursi}, {Cooper}, {Cornez}, {Cowell}, {Crifo}, {Cropper}, {Crosta},
  {Crowley}, {Dafonte}, {Dapergolas}, {David}, {David}, {de Laverny}, {De
  Luise}, {De March}, {De Ridder}, {de Souza}, {de Torres}, {del Peloso}, {del
  Pozo}, {Delbo}, {Delgado}, {Delisle}, {Demouchy}, {Dharmawardena}, {Di
  Matteo}, {Diakite}, {Diener}, {Distefano}, {Dolding}, {Edvardsson}, {Enke},
  {Fabre}, {Fabrizio}, {Faigler}, {Fedorets}, {Fernique}, {Fienga}, {Figueras},
  {Fournier}, {Fouron}, {Fragkoudi}, {Gai}, {Garcia-Gutierrez},
  {Garcia-Reinaldos}, {Garc{\'\i}a-Torres}, {Garofalo}, {Gavel}, {Gavras},
  {Gerlach}, {Geyer}, {Giacobbe}, {Gilmore}, {Girona}, {Giuffrida}, {Gomel},
  {Gomez}, {Gonz{\'a}lez-N{\'u}{\~n}ez}, {Gonz{\'a}lez-Santamar{\'\i}a},
  {Gonz{\'a}lez-Vidal}, {Granvik}, {Guillout}, {Guiraud},
  {Guti{\'e}rrez-S{\'a}nchez}, {Guy}, {Hatzidimitriou}, {Hauser}, {Haywood},
  {Helmer}, {Helmi}, {Sarmiento}, {Hidalgo}, {Hilger}, {H{\l}adczuk}, {Hobbs},
  {Holland}, {Huckle}, {Jardine}, {Jasniewicz}, {Jean-Antoine Piccolo},
  {Jim{\'e}nez-Arranz}, {Jorissen}, {Juaristi Campillo}, {Julbe}, {Karbevska},
  {Kervella}, {Khanna}, {Kontizas}, {Kordopatis}, {Korn}, {K{\'o}sp{\'a}l},
  {Kostrzewa-Rutkowska}, {Kruszy{\'n}ska}, {Kun}, {Laizeau}, {Lambert},
  {Lanza}, {Lasne}, {Le Campion}, {Lebreton}, {Lebzelter}, {Leccia}, {Leclerc},
  {Lecoeur-Taibi}, {Liao}, {Licata}, {Lindstr{\o}m}, {Lister}, {Livanou},
  {Lobel}, {Lorca}, {Loup}, {Madrero Pardo}, {Magdaleno Romeo}, {Managau},
  {Mann}, {Manteiga}, {Marchant}, {Marconi}, {Marcos}, {Marcos Santos},
  {Mar{\'\i}n Pina}, {Marinoni}, {Marocco}, {Marshall}, {Polo},
  {Mart{\'\i}n-Fleitas}, {Marton}, {Mary}, {Masip}, {Massari},
  {Mastrobuono-Battisti}, {Mazeh}, {McMillan}, {Messina}, {Michalik}, {Millar},
  {Mints}, {Molina}, {Molinaro}, {Moln{\'a}r}, {Monari}, {Mongui{\'o}},
  {Montegriffo}, {Montero}, {Mor}, {Mora}, {Morbidelli}, {Morel}, {Morris},
  {Muraveva}, {Murphy}, {Musella}, {Nagy}, {Noval}, {Oca{\~n}a}, {Ogden},
  {Ordenovic}, {Osinde}, {Pagani}, {Pagano}, {Palaversa}, {Palicio},
  {Pallas-Quintela}, {Panahi}, {Payne-Wardenaar}, {Pe{\~n}alosa Esteller},
  {Penttil{\"a}}, {Pichon}, {Piersimoni}, {Pineau}, {Plachy}, {Plum}, {Poggio},
  {Pr{\v{s}}a}, {Pulone}, {Racero}, {Ragaini}, {Rainer}, {Raiteri}, {Rambaux},
  {Ramos}, {Ramos-Lerate}, {Re Fiorentin}, {Regibo}, {Richards}, {Rios Diaz},
  {Ripepi}, {Riva}, {Rix}, {Rixon}, {Robichon}, {Robin}, {Robin}, {Roelens},
  {Rogues}, {Rohrbasser}, {Romero-G{\'o}mez}, {Rowell}, {Royer}, {Ruz Mieres},
  {Rybicki}, {Sadowski}, {S{\'a}ez N{\'u}{\~n}ez}, {Sagrist{\`a} Sell{\'e}s},
  {Sahlmann}, {Salguero}, {Samaras}, {Sanchez Gimenez}, {Sanna},
  {Santove{\~n}a}, {Sarasso}, {Schultheis}, {Sciacca}, {Segol}, {Segovia},
  {S{\'e}gransan}, {Semeux}, {Shahaf}, {Siddiqui}, {Siebert}, {Siltala},
  {Silvelo}, {Slezak}, {Slezak}, {Smart}, {Snaith}, {Solano}, {Solitro},
  {Souami}, {Souchay}, {Spagna}, {Spina}, {Spoto}, {Steele},
  {Steidelm{\"u}ller}, {Stephenson}, {S{\"u}veges}, {Surdej}, {Szabados},
  {Szegedi-Elek}, {Taris}, {Taylo}, {Teixeira}, {Tolomei}, {Tonello}, {Torra},
  {Torra}, {Torralba Elipe}, {Trabucchi}, {Tsounis}, {Turon}, {Ulla}, {Unger},
  {Vaillant}, {van Dillen}, {van Reeven}, {Vanel}, {Vecchiato}, {Viala},
  {Vicente}, {Voutsinas}, {Weiler}, {Wevers}, {Wyrzykowski}, {Yoldas}, {Yvard},
  {Zhao}, {Zorec}, {Zucker}, \& {Zwitter}}]{Gaia2022arXiv220800211G}
{Gaia Collaboration}, {Vallenari}, A., {Brown}, A.~G.~A., {et~al.} 2022, arXiv
  e-prints, arXiv:2208.00211

\bibitem[{{Gloudemans} {et~al.}(2022){Gloudemans}, {Duncan}, {Saxena},
  {Harikane}, {Hill}, {Zeimann}, {R{\"o}ttgering}, {Yang}, {Best},
  {Ba{\~n}ados}, {Drabent}, {Hardcastle}, {Hennawi}, {Lansbury},
  {Magliocchetti}, {Miley}, {Nanni}, {Shimwell}, {Smith}, {Venemans}, \&
  {Wagenveld}}]{Gloudemans2022A&A...668A..27G}
{Gloudemans}, A.~J., {Duncan}, K.~J., {Saxena}, A., {et~al.} 2022, \aap, 668,
  A27

\bibitem[{{Hallinan} {et~al.}(2007){Hallinan}, {Bourke}, {Lane}, {Antonova},
  {Zavala}, {Brisken}, {Boyle}, {Vrba}, {Doyle}, \&
  {Golden}}]{Hallinan2007ApJ...663L..25H}
{Hallinan}, G., {Bourke}, S., {Lane}, C., {et~al.} 2007, \apjl, 663, L25

\bibitem[{{Hallinan} {et~al.}(2015){Hallinan}, {Littlefair}, {Cotter},
  {Bourke}, {Harding}, {Pineda}, {Butler}, {Golden}, {Basri}, {Doyle}, {Kao},
  {Berdyugina}, {Kuznetsov}, {Rupen}, \&
  {Antonova}}]{Hallinan2015Natur.523..568H}
{Hallinan}, G., {Littlefair}, S.~P., {Cotter}, G., {et~al.} 2015, \nat, 523,
  568

\bibitem[{{Hewett} {et~al.}(2006){Hewett}, {Warren}, {Leggett}, \&
  {Hodgkin}}]{Hewett2006MNRAS.367..454H}
{Hewett}, P.~C., {Warren}, S.~J., {Leggett}, S.~K., \& {Hodgkin}, S.~T. 2006,
  \mnras, 367, 454

\bibitem[{{Hill} {et~al.}(2021){Hill}, {Lee}, {MacQueen}, {Kelz}, {Drory},
  {Vattiat}, {Good}, {Ramsey}, {Kriel}, {Peterson}, {DePoy}, {Gebhardt},
  {Marshall}, {Tuttle}, {Bauer}, {Chonis}, {Fabricius}, {Froning},
  {H{\"a}user}, {Indahl}, {Jahn}, {Landriau}, {Leck}, {Montesano}, {Prochaska},
  {Snigula}, {Zeimann}, {Bryant}, {Damm}, {Fowler}, {Janowiecki}, {Martin},
  {Mrozinski}, {Odewahn}, {Rostopchin}, {Shetrone}, {Spencer}, {Mentuch
  Cooper}, {Armandroff}, {Bender}, {Dalton}, {Hopp}, {Komatsu}, {Nicklas},
  {Ramsey}, {Roth}, {Schneider}, {Sneden}, \&
  {Steinmetz}}]{Hill2021AJ....162..298H}
{Hill}, G.~J., {Lee}, H., {MacQueen}, P.~J., {et~al.} 2021, \aj, 162, 298

\bibitem[{{Kao} {et~al.}(2016){Kao}, {Hallinan}, {Pineda}, {Escala},
  {Burgasser}, {Bourke}, \& {Stevenson}}]{Kao2016ApJ...818...24K}
{Kao}, M.~M., {Hallinan}, G., {Pineda}, J.~S., {et~al.} 2016, \apj, 818, 24

\bibitem[{{Kao} {et~al.}(2018){Kao}, {Hallinan}, {Pineda}, {Stevenson}, \&
  {Burgasser}}]{Kao2018ApJS..237...25K}
{Kao}, M.~M., {Hallinan}, G., {Pineda}, J.~S., {Stevenson}, D., \& {Burgasser},
  A. 2018, \apjs, 237, 25

\bibitem[{{Kashikawa} {et~al.}(2002){Kashikawa}, {Aoki}, {Asai}, {Ebizuka},
  {Inata}, {Iye}, {Kawabata}, {Kosugi}, {Ohyama}, {Okita}, {Ozawa}, {Saito},
  {Sasaki}, {Sekiguchi}, {Shimizu}, {Taguchi}, {Takata}, {Yadoumaru}, \&
  {Yoshida}}]{Kashikawa2002PASJ...54..819K}
{Kashikawa}, N., {Aoki}, K., {Asai}, R., {et~al.} 2002, \pasj, 54, 819

\bibitem[{{Kondapally} {et~al.}(2021){Kondapally}, {Best}, {Hardcastle},
  {Nisbet}, {Bonato}, {Sabater}, {Duncan}, {McCheyne}, {Cochrane}, {Bowler},
  {Williams}, {Shimwell}, {Tasse}, {Croston}, {Goyal}, {Jamrozy}, {Jarvis},
  {Mahatma}, {R{\"o}ttgering}, {Smith}, {Wo{\l}owska}, {Bondi}, {Brienza},
  {Brown}, {Br{\"u}ggen}, {Chambers}, {Garrett}, {G{\"u}rkan}, {Huber},
  {Kunert-Bajraszewska}, {Magnier}, {Mingo}, {Mostert},
  {Nikiel-Wroczy{\'n}ski}, {O'Sullivan}, {Paladino}, {Ploeckinger}, {Prandoni},
  {Rosenthal}, {Schwarz}, {Shulevski}, {Wagenveld}, \&
  {Wang}}]{Kondapally2021A&A...648A...3K}
{Kondapally}, R., {Best}, P.~N., {Hardcastle}, M.~J., {et~al.} 2021, \aap, 648,
  A3

\bibitem[{{Lacy} {et~al.}(2020){Lacy}, {Baum}, {Chandler}, {Chatterjee},
  {Clarke}, {Deustua}, {English}, {Farnes}, {Gaensler}, {Gugliucci},
  {Hallinan}, {Kent}, {Kimball}, {Law}, {Lazio}, {Marvil}, {Mao}, {Medlin},
  {Mooley}, {Murphy}, {Myers}, {Osten}, {Richards}, {Rosolowsky}, {Rudnick},
  {Schinzel}, {Sivakoff}, {Sjouwerman}, {Taylor}, {White}, {Wrobel},
  {Andernach}, {Beasley}, {Berger}, {Bhatnager}, {Birkinshaw}, {Bower},
  {Brandt}, {Brown}, {Burke-Spolaor}, {Butler}, {Comerford}, {Demorest}, {Fu},
  {Giacintucci}, {Golap}, {G{\"u}th}, {Hales}, {Hiriart}, {Hodge}, {Horesh},
  {Ivezi{\'c}}, {Jarvis}, {Kamble}, {Kassim}, {Liu}, {Loinard}, {Lyons},
  {Masters}, {Mezcua}, {Moellenbrock}, {Mroczkowski}, {Nyland}, {O'Dea},
  {O'Sullivan}, {Peters}, {Radford}, {Rao}, {Robnett}, {Salcido}, {Shen},
  {Sobotka}, {Witz}, {Vaccari}, {van Weeren}, {Vargas}, {Williams}, \&
  {Yoon}}]{Lacy2020PASP..132c5001L}
{Lacy}, M., {Baum}, S.~A., {Chandler}, C.~J., {et~al.} 2020, \pasp, 132, 035001

\bibitem[{{Lawrence} {et~al.}(2007){Lawrence}, {Warren}, {Almaini}, {Edge},
  {Hambly}, {Jameson}, {Lucas}, {Casali}, {Adamson}, {Dye}, {Emerson},
  {Foucaud}, {Hewett}, {Hirst}, {Hodgkin}, {Irwin}, {Lodieu}, {McMahon},
  {Simpson}, {Smail}, {Mortlock}, \& {Folger}}]{Lawrence2007MNRAS.379.1599L}
{Lawrence}, A., {Warren}, S.~J., {Almaini}, O., {et~al.} 2007, \mnras, 379,
  1599

\bibitem[{{Lynch} {et~al.}(2017){Lynch}, {Lenc}, {Kaplan}, {Murphy}, \&
  {Anderson}}]{Lynch2017ApJ...836L..30L}
{Lynch}, C.~R., {Lenc}, E., {Kaplan}, D.~L., {Murphy}, T., \& {Anderson}, G.~E.
  2017, \apjl, 836, L30

\bibitem[{{Oke} \& {Gunn}(1983)}]{Oke1983ApJ...266..713O}
{Oke}, J.~B. \& {Gunn}, J.~E. 1983, \apj, 266, 713

\bibitem[{{Osten} {et~al.}(2005){Osten}, {Hawley}, {Allred}, {Johns-Krull}, \&
  {Roark}}]{2005ApJ...621..398O}
{Osten}, R.~A., {Hawley}, S.~L., {Allred}, J.~C., {Johns-Krull}, C.~M., \&
  {Roark}, C. 2005, \apj, 621, 398

\bibitem[{{Pineda} {et~al.}(2017){Pineda}, {Hallinan}, \&
  {Kao}}]{Pineda2017ApJ...846...75P}
{Pineda}, J.~S., {Hallinan}, G., \& {Kao}, M.~M. 2017, \apj, 846, 75

\bibitem[{{Ramsey} {et~al.}(1998){Ramsey}, {Engel}, {Rhoads}, {Maywalt},
  {McGouldrick}, \& {Andersen}}]{Ramsey1998AAS...193.1007R}
{Ramsey}, L.~W., {Engel}, L., {Rhoads}, B., {et~al.} 1998, in American
  Astronomical Society Meeting Abstracts, Vol. 193, American Astronomical
  Society Meeting Abstracts, 10.07

\bibitem[{{Reiners} \& {Basri}(2010)}]{Reiners2010ApJ...710..924R}
{Reiners}, A. \& {Basri}, G. 2010, \apj, 710, 924

\bibitem[{{Reiners} \& {Christensen}(2010)}]{reiners2010A&A...522A..13R}
{Reiners}, A. \& {Christensen}, U.~R. 2010, \aap, 522, A13

\bibitem[{{Roulston} {et~al.}(2020){Roulston}, {Green}, \&
  {Kesseli}}]{Roulston2020ApJS..249...34R}
{Roulston}, B.~R., {Green}, P.~J., \& {Kesseli}, A.~Y. 2020, \apjs, 249, 34

\bibitem[{{Route} \& {Wolszczan}(2012)}]{Route2012ApJ...747L..22R}
{Route}, M. \& {Wolszczan}, A. 2012, \apjl, 747, L22

\bibitem[{{Route} \& {Wolszczan}(2016)}]{Route2016ApJ...830...85R}
{Route}, M. \& {Wolszczan}, A. 2016, \apj, 830, 85

\bibitem[{{Sabater} {et~al.}(2021){Sabater}, {Best}, {Tasse}, {Hardcastle},
  {Shimwell}, {Nisbet}, {Jelic}, {Callingham}, {R{\"o}ttgering}, {Bonato},
  {Bondi}, {Ciardi}, {Cochrane}, {Jarvis}, {Kondapally}, {Koopmans},
  {O'Sullivan}, {Prandoni}, {Schwarz}, {Smith}, {Wang}, {Williams}, \&
  {Zaroubi}}]{sabater2021A&A...648A...2S}
{Sabater}, J., {Best}, P.~N., {Tasse}, C., {et~al.} 2021, \aap, 648, A2

\bibitem[{{Shimwell} {et~al.}(2022){Shimwell}, {Hardcastle}, {Tasse}, {Best},
  {R{\"o}ttgering}, {Williams}, {Botteon}, {Drabent}, {Mechev}, {Shulevski},
  {van Weeren}, {Bester}, {Br{\"u}ggen}, {Brunetti}, {Callingham}, {Chy{\.z}y},
  {Conway}, {Dijkema}, {Duncan}, {de Gasperin}, {Hale}, {Haverkorn}, {Hugo},
  {Jackson}, {Mevius}, {Miley}, {Morabito}, {Morganti}, {Offringa}, {Oonk},
  {Rafferty}, {Sabater}, {Smith}, {Schwarz}, {Smirnov}, {O'Sullivan},
  {Vedantham}, {White}, {Albert}, {Alegre}, {Asabere}, {Bacon}, {Bonafede},
  {Bonnassieux}, {Brienza}, {Bilicki}, {Bonato}, {Calistro Rivera}, {Cassano},
  {Cochrane}, {Croston}, {Cuciti}, {Dallacasa}, {Danezi}, {Dettmar}, {Di
  Gennaro}, {Edler}, {En{\ss}lin}, {Emig}, {Franzen}, {Garc{\'\i}a-Vergara},
  {Grange}, {G{\"u}rkan}, {Hajduk}, {Heald}, {Heesen}, {Hoang}, {Hoeft},
  {Horellou}, {Iacobelli}, {Jamrozy}, {Jeli{\'c}}, {Kondapally}, {Kukreti},
  {Kunert-Bajraszewska}, {Magliocchetti}, {Mahatma}, {Ma{\l}ek}, {Mandal},
  {Massaro}, {Meyer-Zhao}, {Mingo}, {Mostert}, {Nair}, {Nakoneczny},
  {Nikiel-Wroczy{\'n}ski}, {Orr{\'u}}, {Pajdosz-{\'S}mierciak}, {Pasini},
  {Prandoni}, {van Piggelen}, {Rajpurohit}, {Retana-Montenegro}, {Riseley},
  {Rowlinson}, {Saxena}, {Schrijvers}, {Sweijen}, {Siewert}, {Timmerman},
  {Vaccari}, {Vink}, {West}, {Wo{\l}owska}, {Zhang}, \&
  {Zheng}}]{Shimwell2022A&A...659A...1S}
{Shimwell}, T.~W., {Hardcastle}, M.~J., {Tasse}, C., {et~al.} 2022, \aap, 659,
  A1

\bibitem[{{Skrutskie} {et~al.}(2006){Skrutskie}, {Cutri}, {Stiening},
  {Weinberg}, {Schneider}, {Carpenter}, {Beichman}, {Capps}, {Chester},
  {Elias}, {Huchra}, {Liebert}, {Lonsdale}, {Monet}, {Price}, {Seitzer},
  {Jarrett}, {Kirkpatrick}, {Gizis}, {Howard}, {Evans}, {Fowler}, {Fullmer},
  {Hurt}, {Light}, {Kopan}, {Marsh}, {McCallon}, {Tam}, {Van Dyk}, \&
  {Wheelock}}]{Skrutskie2006AJ....131.1163S}
{Skrutskie}, M.~F., {Cutri}, R.~M., {Stiening}, R., {et~al.} 2006, \aj, 131,
  1163

\bibitem[{{Sutherland} \& {Saunders}(1992)}]{Sutherland1992MNRAS.259..413S}
{Sutherland}, W. \& {Saunders}, W. 1992, \mnras, 259, 413

\bibitem[{{Tasse} {et~al.}(2021){Tasse}, {Shimwell}, {Hardcastle},
  {O'Sullivan}, {van Weeren}, {Best}, {Bester}, {Hugo}, {Smirnov}, {Sabater},
  {Calistro-Rivera}, {de Gasperin}, {Morabito}, {R{\"o}ttgering}, {Williams},
  {Bonato}, {Bondi}, {Botteon}, {Br{\"u}ggen}, {Brunetti}, {Chy{\.z}y},
  {Garrett}, {G{\"u}rkan}, {Jarvis}, {Kondapally}, {Mandal}, {Prandoni},
  {Repetti}, {Retana-Montenegro}, {Schwarz}, {Shulevski}, \&
  {Wiaux}}]{tasse2021A&A...648A...1T}
{Tasse}, C., {Shimwell}, T., {Hardcastle}, M.~J., {et~al.} 2021, \aap, 648, A1

\bibitem[{{Tingay} {et~al.}(2013){Tingay}, {Goeke}, {Bowman}, {Emrich}, {Ord},
  {Mitchell}, {Morales}, {Booler}, {Crosse}, {Wayth}, {Lonsdale}, {Tremblay},
  {Pallot}, {Colegate}, {Wicenec}, {Kudryavtseva}, {Arcus}, {Barnes},
  {Bernardi}, {Briggs}, {Burns}, {Bunton}, {Cappallo}, {Corey}, {Deshpande},
  {Desouza}, {Gaensler}, {Greenhill}, {Hall}, {Hazelton}, {Herne}, {Hewitt},
  {Johnston-Hollitt}, {Kaplan}, {Kasper}, {Kincaid}, {Koenig}, {Kratzenberg},
  {Lynch}, {Mckinley}, {Mcwhirter}, {Morgan}, {Oberoi}, {Pathikulangara},
  {Prabu}, {Remillard}, {Rogers}, {Roshi}, {Salah}, {Sault}, {Udaya-Shankar},
  {Schlagenhaufer}, {Srivani}, {Stevens}, {Subrahmanyan}, {Waterson},
  {Webster}, {Whitney}, {Williams}, {Williams}, \&
  {Wyithe}}]{Tingay2013PASA...30....7T}
{Tingay}, S.~J., {Goeke}, R., {Bowman}, J.~D., {et~al.} 2013, \pasa, 30, e007

\bibitem[{{Tody}(1986)}]{Tody1986SPIE..627..733T}
{Tody}, D. 1986, in Society of Photo-Optical Instrumentation Engineers (SPIE)
  Conference Series, Vol. 627, Instrumentation in astronomy VI, ed. D.~L.
  {Crawford}, 733

\bibitem[{{Toet} {et~al.}(2021){Toet}, {Vedantham}, {Callingham}, {Veken},
  {Shimwell}, {Zarka}, {R{\"o}ttgering}, \& {Drabent}}]{2021A&A...654A..21T}
{Toet}, S.~E.~B., {Vedantham}, H.~K., {Callingham}, J.~R., {et~al.} 2021, \aap,
  654, A21

\bibitem[{{Trigilio} {et~al.}(2023){Trigilio}, {Biswas}, {Leto}, {Umana},
  {Busa}, {Cavallaro}, {Das}, {Chandra}, {Perez-Torres}, {Wade}, {Bordiu},
  {Buemi}, {Bufano}, {Ingallinera}, {Loru}, \&
  {Riggi}}]{Trigilio2023arXiv230500809T}
{Trigilio}, C., {Biswas}, A., {Leto}, P., {et~al.} 2023, arXiv e-prints,
  arXiv:2305.00809

\bibitem[{{van Haarlem} {et~al.}(2013){van Haarlem}, {Wise}, {Gunst}, {Heald},
  {McKean}, {Hessels}, {de Bruyn}, {Nijboer}, {Swinbank}, {Fallows},
  {Brentjens}, {Nelles}, {Beck}, {Falcke}, {Fender}, {H{\"o}randel},
  {Koopmans}, {Mann}, {Miley}, {R{\"o}ttgering}, {Stappers}, {Wijers},
  {Zaroubi}, {van den Akker}, {Alexov}, {Anderson}, {Anderson}, {van Ardenne},
  {Arts}, {Asgekar}, {Avruch}, {Batejat}, {B{\"a}hren}, {Bell}, {Bell}, {van
  Bemmel}, {Bennema}, {Bentum}, {Bernardi}, {Best}, {B{\^\i}rzan}, {Bonafede},
  {Boonstra}, {Braun}, {Bregman}, {Breitling}, {van de Brink}, {Broderick},
  {Broekema}, {Brouw}, {Br{\"u}ggen}, {Butcher}, {van Cappellen}, {Ciardi},
  {Coenen}, {Conway}, {Coolen}, {Corstanje}, {Damstra}, {Davies}, {Deller},
  {Dettmar}, {van Diepen}, {Dijkstra}, {Donker}, {Doorduin}, {Dromer}, {Drost},
  {van Duin}, {Eisl{\"o}ffel}, {van Enst}, {Ferrari}, {Frieswijk}, {Gankema},
  {Garrett}, {de Gasperin}, {Gerbers}, {de Geus}, {Grie{\ss}meier}, {Grit},
  {Gruppen}, {Hamaker}, {Hassall}, {Hoeft}, {Holties}, {Horneffer}, {van der
  Horst}, {van Houwelingen}, {Huijgen}, {Iacobelli}, {Intema}, {Jackson},
  {Jelic}, {de Jong}, {Juette}, {Kant}, {Karastergiou}, {Koers}, {Kollen},
  {Kondratiev}, {Kooistra}, {Koopman}, {Koster}, {Kuniyoshi}, {Kramer},
  {Kuper}, {Lambropoulos}, {Law}, {van Leeuwen}, {Lemaitre}, {Loose}, {Maat},
  {Macario}, {Markoff}, {Masters}, {McFadden}, {McKay-Bukowski}, {Meijering},
  {Meulman}, {Mevius}, {Middelberg}, {Millenaar}, {Miller-Jones}, {Mohan},
  {Mol}, {Morawietz}, {Morganti}, {Mulcahy}, {Mulder}, {Munk}, {Nieuwenhuis},
  {van Nieuwpoort}, {Noordam}, {Norden}, {Noutsos}, {Offringa}, {Olofsson},
  {Omar}, {Orr{\'u}}, {Overeem}, {Paas}, {Pand ey-Pommier}, {Pandey}, {Pizzo},
  {Polatidis}, {Rafferty}, {Rawlings}, {Reich}, {de Reijer}, {Reitsma},
  {Renting}, {Riemers}, {Rol}, {Romein}, {Roosjen}, {Ruiter}, {Scaife}, {van
  der Schaaf}, {Scheers}, {Schellart}, {Schoenmakers}, {Schoonderbeek},
  {Serylak}, {Shulevski}, {Sluman}, {Smirnov}, {Sobey}, {Spreeuw}, {Steinmetz},
  {Sterks}, {Stiepel}, {Stuurwold}, {Tagger}, {Tang}, {Tasse}, {Thomas},
  {Thoudam}, {Toribio}, {van der Tol}, {Usov}, {van Veelen}, {van der Veen},
  {ter Veen}, {Verbiest}, {Vermeulen}, {Vermaas}, {Vocks}, {Vogt}, {de Vos},
  {van der Wal}, {van Weeren}, {Weggemans}, {Weltevrede}, {White}, {Wijnholds},
  {Wilhelmsson}, {Wucknitz}, {Yatawatta}, {Zarka}, {Zensus}, \& {van
  Zwieten}}]{vanHaarlem2013A&A...556A...2V}
{van Haarlem}, M.~P., {Wise}, M.~W., {Gunst}, A.~W., {et~al.} 2013, \aap, 556,
  A2

\bibitem[{{Vedantham} {et~al.}(2022){Vedantham}, {Callingham}, {Shimwell},
  {Benz}, {Hajduk}, {Ray}, {Tasse}, \& {Drabent}}]{2022ApJ...926L..30V}
{Vedantham}, H.~K., {Callingham}, J.~R., {Shimwell}, T.~W., {et~al.} 2022,
  \apjl, 926, L30

\bibitem[{{Vedantham} {et~al.}(2020){Vedantham}, {Callingham}, {Shimwell},
  {Dupuy}, {Best}, {Liu}, {Zhang}, {De}, {Lamy}, {Zarka}, {R{\"o}ttgering}, \&
  {Shulevski}}]{Vedantham2020ApJ...903L..33V}
{Vedantham}, H.~K., {Callingham}, J.~R., {Shimwell}, T.~W., {et~al.} 2020,
  \apjl, 903, L33

\bibitem[{{Vedantham} {et~al.}(2023){Vedantham}, {Dupuy}, {Evans}, {Sanghi},
  {Callingham}, {Shimwell}, {Best}, {Liu}, \&
  {Zarka}}]{Vedantham2023A&A...675L...6V}
{Vedantham}, H.~K., {Dupuy}, T.~J., {Evans}, E.~L., {et~al.} 2023, \aap, 675,
  L6

\bibitem[{{Villadsen} \& {Hallinan}(2019)}]{2019ApJ...871..214V}
{Villadsen}, J. \& {Hallinan}, G. 2019, \apj, 871, 214

\bibitem[{{Wagenveld} {et~al.}(2022){Wagenveld}, {Saxena}, {Duncan},
  {R{\"o}ttgering}, \& {Zhang}}]{Wagenveld2022A&A...660A..22W}
{Wagenveld}, J.~D., {Saxena}, A., {Duncan}, K.~J., {R{\"o}ttgering}, H.~J.~A.,
  \& {Zhang}, M. 2022, \aap, 660, A22

\bibitem[{{West} {et~al.}(2011){West}, {Morgan}, {Bochanski}, {Andersen},
  {Bell}, {Kowalski}, {Davenport}, {Hawley}, {Schmidt}, {Bernat}, {Hilton},
  {Muirhead}, {Covey}, {Rojas-Ayala}, {Schlawin}, {Gooding}, {Schluns},
  {Dhital}, {Pineda}, \& {Jones}}]{west2011ASPC..448.1407W}
{West}, A.~A., {Morgan}, D.~P., {Bochanski}, J.~J., {et~al.} 2011, in
  Astronomical Society of the Pacific Conference Series, Vol. 448, 16th
  Cambridge Workshop on Cool Stars, Stellar Systems, and the Sun, ed.
  C.~{Johns-Krull}, M.~K. {Browning}, \& A.~A. {West}, 1407

\bibitem[{{Wilson} {et~al.}(2009){Wilson}, {Muzzin}, {Yee}, {Lacy}, {Surace},
  {Gilbank}, {Blindert}, {Hoekstra}, {Majumdar}, {Demarco}, {Gardner},
  {Gladders}, \& {Lonsdale}}]{Wilson2009ApJ...698.1943W}
{Wilson}, G., {Muzzin}, A., {Yee}, H.~K.~C., {et~al.} 2009, \apj, 698, 1943

\bibitem[{{Wright} {et~al.}(2010){Wright}, {Eisenhardt}, {Mainzer}, {Ressler},
  {Cutri}, {Jarrett}, {Kirkpatrick}, {Padgett}, {McMillan}, {Skrutskie},
  {Stanford}, {Cohen}, {Walker}, {Mather}, {Leisawitz}, {Gautier}, {McLean},
  {Benford}, {Lonsdale}, {Blain}, {Mendez}, {Irace}, {Duval}, {Liu}, {Royer},
  {Heinrichsen}, {Howard}, {Shannon}, {Kendall}, {Walsh}, {Larsen}, {Cardon},
  {Schick}, {Schwalm}, {Abid}, {Fabinsky}, {Naes}, \&
  {Tsai}}]{wright2010AJ....140.1868W}
{Wright}, E.~L., {Eisenhardt}, P. R.~M., {Mainzer}, A.~K., {et~al.} 2010, \aj,
  140, 1868

\bibitem[{{York} {et~al.}(2000){York}, {Adelman}, {Anderson}, {Anderson},
  {Annis}, {Bahcall}, {Bakken}, {Barkhouser}, {Bastian}, {Berman}, {Boroski},
  {Bracker}, {Briegel}, {Briggs}, {Brinkmann}, {Brunner}, {Burles}, {Carey},
  {Carr}, {Castander}, {Chen}, {Colestock}, {Connolly}, {Crocker}, {Csabai},
  {Czarapata}, {Davis}, {Doi}, {Dombeck}, {Eisenstein}, {Ellman}, {Elms},
  {Evans}, {Fan}, {Federwitz}, {Fiscelli}, {Friedman}, {Frieman}, {Fukugita},
  {Gillespie}, {Gunn}, {Gurbani}, {de Haas}, {Haldeman}, {Harris}, {Hayes},
  {Heckman}, {Hennessy}, {Hindsley}, {Holm}, {Holmgren}, {Huang}, {Hull},
  {Husby}, {Ichikawa}, {Ichikawa}, {Ivezi{\'c}}, {Kent}, {Kim}, {Kinney},
  {Klaene}, {Kleinman}, {Kleinman}, {Knapp}, {Korienek}, {Kron}, {Kunszt},
  {Lamb}, {Lee}, {Leger}, {Limmongkol}, {Lindenmeyer}, {Long}, {Loomis},
  {Loveday}, {Lucinio}, {Lupton}, {MacKinnon}, {Mannery}, {Mantsch}, {Margon},
  {McGehee}, {McKay}, {Meiksin}, {Merelli}, {Monet}, {Munn}, {Narayanan},
  {Nash}, {Neilsen}, {Neswold}, {Newberg}, {Nichol}, {Nicinski}, {Nonino},
  {Okada}, {Okamura}, {Ostriker}, {Owen}, {Pauls}, {Peoples}, {Peterson},
  {Petravick}, {Pier}, {Pope}, {Pordes}, {Prosapio}, {Rechenmacher}, {Quinn},
  {Richards}, {Richmond}, {Rivetta}, {Rockosi}, {Ruthmansdorfer}, {Sandford},
  {Schlegel}, {Schneider}, {Sekiguchi}, {Sergey}, {Shimasaku}, {Siegmund},
  {Smee}, {Smith}, {Snedden}, {Stone}, {Stoughton}, {Strauss}, {Stubbs},
  {SubbaRao}, {Szalay}, {Szapudi}, {Szokoly}, {Thakar}, {Tremonti}, {Tucker},
  {Uomoto}, {Vanden Berk}, {Vogeley}, {Waddell}, {Wang}, {Watanabe},
  {Weinberg}, {Yanny}, {Yasuda}, \& {SDSS
  Collaboration}}]{York2000AJ....120.1579Y}
{York}, D.~G., {Adelman}, J., {Anderson}, John~E., J., {et~al.} 2000, \aj, 120,
  1579

\bibitem[{{Zarka}(1998)}]{zarka1998}
{Zarka}, P. 1998, \jgr, 103, 20159

\bibitem[{{Zic} {et~al.}(2019){Zic}, {Lynch}, {Murphy}, {Kaplan}, \&
  {Chandra}}]{2019MNRAS.483..614Z}
{Zic}, A., {Lynch}, C., {Murphy}, T., {Kaplan}, D.~L., \& {Chandra}, P. 2019,
  \mnras, 483, 614

\end{thebibliography}

\begin{appendix}

\section{Cutouts of M dwarf sample}
\label{sec:appendix_cutouts}

\begin{figure*}[p]
\makebox[\linewidth]{
        \includegraphics[width=0.8\textwidth, trim={0.0cm 2.15cm 0cm 0.0cm}, clip]{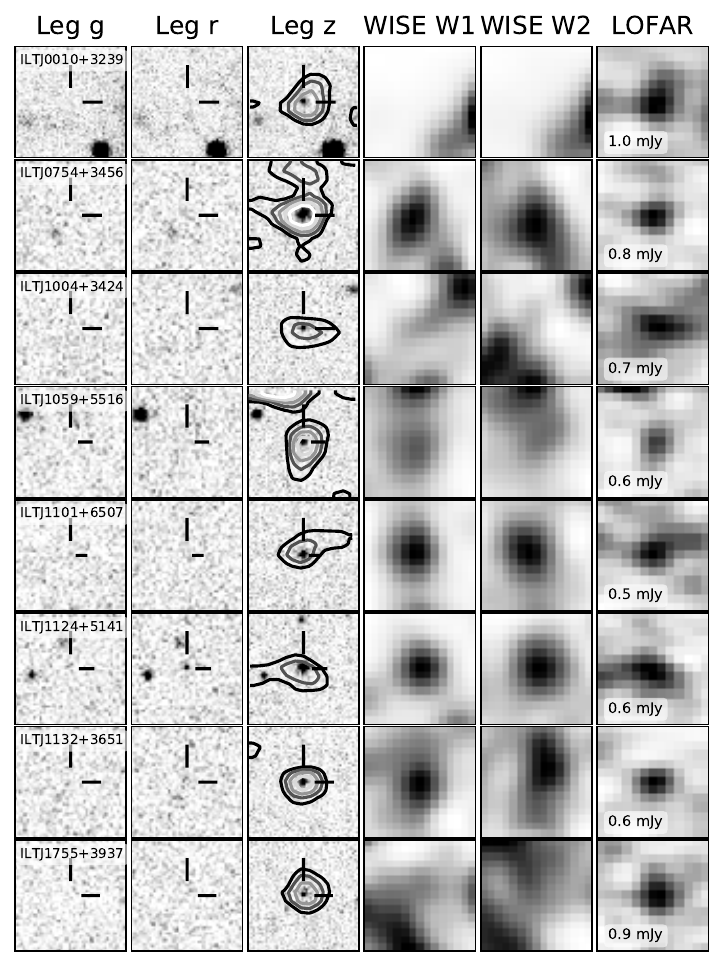}
    }
     
     \caption{20"$\times$20" Legacy DR8 \citep{dey2019AJ....157..168D}, AllWISE \citep{Cutri2014yCat.2328....0C}, and LOFAR (LoTSS-DR2; \citealt{Shimwell2022A&A...659A...1S}) cutouts for our brown dwarf sample. Radio contours are drawn at 2-6 times the rms noise level.}
     \label{fig:cutouts}
\end{figure*}

The optical, infrared, and radio cutouts of our M\,dwarfs are shown in Fig.~\ref{fig:cutouts}. They have been selected as high-$z$ quasar candidates by \cite{Gloudemans2022A&A...668A..27G} and are non-detected in both the Legacy $g$ and $r$ band.

\section{Rejected candidates}
\label{sec:appendix_rejected}

The rejected M\,dwarfs are summarised in Table~\ref{tab:sample_props_rejected} including their optical coordinates, radio source separation, and $r$ statistic value. In this paper, we conservatively retain only the M\,dwarfs within $r < 2$ and radio-optical source separations of $< 1\arcsec$, as the most closely matched associations with best confidence. 

\begin{table}
\caption{Properties of the rejected candidates. The separation reported is the offset between the optical and radio source positions from the Legacy Surveys and LoTSS-DR2 catalogue, respectively.}
\label{tab:sample_props_rejected}
\centering
\resizebox{0.95\columnwidth}{!}{
\begin{tabular}{c c c c}
\hline\hline
Source name & Optical coordinates  & Separation & $r$ \\ 
 & (J2000) & ($\arcsec$) &   \\
\hline
ILTJ0729+3809 & 07:29:03 +38:09:45 & 1.1 & 1.8  \\
ILTJ0754+3610 & 07:54:35 +36:10:31 & 1.9 & 7.3  \\
ILTJ0821+4831 & 08:21:47 +48:31:00 & 1.7 & 4.6  \\
ILTJ0909+4502 & 09:09:31 +45:02:01 & 1.7 & 34.1  \\
ILTJ0933+5215 & 09:33:42 +52:15:17 & 1.0 & 5.6  \\
ILTJ0933+5248 & 09:33:59 +52:48:02 & 1.7 & 21.6  \\
ILTJ0950+3440 & 09:50:21 +34:40:29 & 1.0 & 4.6  \\
ILTJ1012+6532 & 10:12:17 +65:32:10 & 1.4 & 8.8  \\
ILTJ1020+5909 & 10:20:01 +59:09:01 & 0.9 & 3.6  \\
ILTJ1056+3112 & 10:56:15 +31:12:32 & 0.6 & 2.6  \\
ILTJ1056+4759 & 10:56:23 +47:59:06 & 0.6 & 2.3  \\
ILTJ1135+3336 & 11:35:04 +33:36:39 & 1.2 & 3.3  \\
ILTJ1354+4234 & 13:54:51 +42:34:01 & 1.3 & 4.3  \\
ILTJ1402+2914 & 14:02:02 +29:14:14 & 1.1 & 6.7  \\
ILTJ2232+1840 & 22:32:16 +18:40:04 & 0.6 & 4.9  \\
ILTJ2235+1802 & 22:35:25 +18:02:37 & 1.2 & 5.0  \\
ILTJ2244+2140 & 22:44:48 +21:40:54 & 1.7 & 4.5  \\
ILTJ2259+2018 & 22:59:57 +20:18:49 & 0.9 & 3.5  \\
ILTJ2316+3142 & 23:16:09 +31:42:40 & 1.4 & 1.6  \\
ILTJ2327+2002 & 23:27:56 +20:02:50 & 1.9 & 3.8  \\
\hline \hline
\end{tabular}}
\end{table}

\end{appendix}

\end{document}